\documentclass[a4paper,twocolumn,11pt,accepted=2023-06-03]{quantumarticle}
\pdfoutput=1
\usepackage[utf8]{inputenc}
\usepackage[english]{babel}
\usepackage[T1]{fontenc}
\usepackage{amsmath, amssymb}
\usepackage{hyperref}
\usepackage[numbers, sort&compress]{natbib}

\usepackage{subcaption}

\usepackage{tikz}

\newtheorem{theorem}{Theorem}

\begin{document}

\title{Quantum Goemans-Williamson Algorithm with the Hadamard Test and Approximate Amplitude Constraints}

\author{Taylor L. Patti}
\email[]{taylorpatti@g.harvard.edu}
\affiliation{Department of Physics, Harvard University, Cambridge, Massachusetts 02138, USA}
\affiliation{NVIDIA, Santa Clara, California 95051, USA}
\author{Jean Kossaifi}
\affiliation{NVIDIA, Santa Clara, California 95051, USA}
\author{Anima Anandkumar}
\affiliation{Department of Computing + Mathematical Sciences (CMS), California Institute of Technology (Caltech), Pasadena, CA 91125 USA}
\affiliation{NVIDIA, Santa Clara, California 95051, USA}
\author{Susanne F. Yelin}
\affiliation{Department of Physics, Harvard University, Cambridge, Massachusetts 02138, USA}

\begin{abstract}
  Semidefinite programs are optimization methods with a wide array of applications, such as approximating difficult combinatorial problems. One such semidefinite program is the Goemans-Williamson algorithm, a popular integer relaxation technique. We introduce a variational quantum algorithm for the Goemans-Williamson algorithm that uses only $n{+}1$ qubits, a constant number of circuit preparations, and $\text{poly}(n)$ expectation values in order to approximately solve semidefinite programs with up to $N=2^n$ variables and $M \sim O(N)$ constraints. Efficient optimization is achieved by encoding the objective matrix as a properly parameterized unitary conditioned on an auxilary qubit, a technique known as the Hadamard Test. The Hadamard Test enables us to optimize the objective function by estimating only a single expectation value of the ancilla qubit, rather than separately estimating exponentially many expectation values. Similarly, we illustrate that the semidefinite programming constraints can be effectively enforced by implementing a second Hadamard Test, as well as imposing a polynomial number of Pauli string amplitude constraints. We demonstrate the effectiveness of our protocol by devising an efficient quantum implementation of the Goemans-Williamson algorithm for various NP-hard problems, including MaxCut. Our method exceeds the performance of analogous classical methods on a diverse subset of well-studied MaxCut problems from the GSet library.
\end{abstract}

\begin{figure*}[ht!]
    \begin{subfigure}{\textwidth}
      \centering
        \includegraphics[height=4cm]{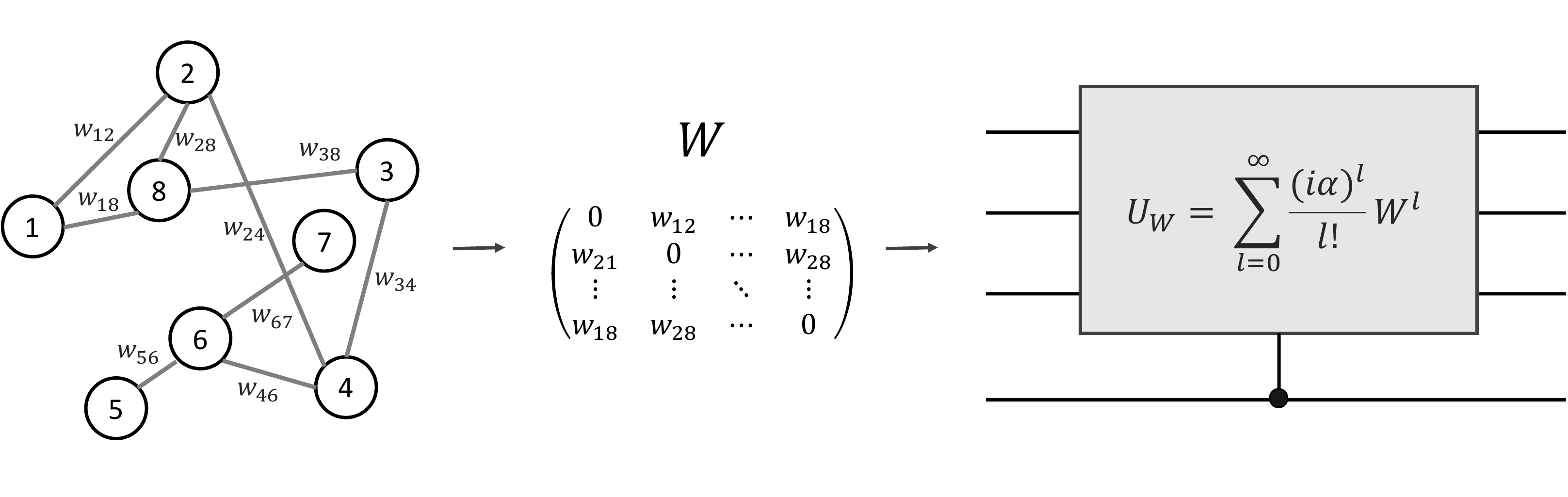}
        \caption{\textbf{Controlled-unitary objective function}}
        \label{fig:main-a}
    \end{subfigure}
    \newline
    \vspace{10pt}
    \begin{subfigure}{\textwidth}
      \centering
        \includegraphics[height=4cm]{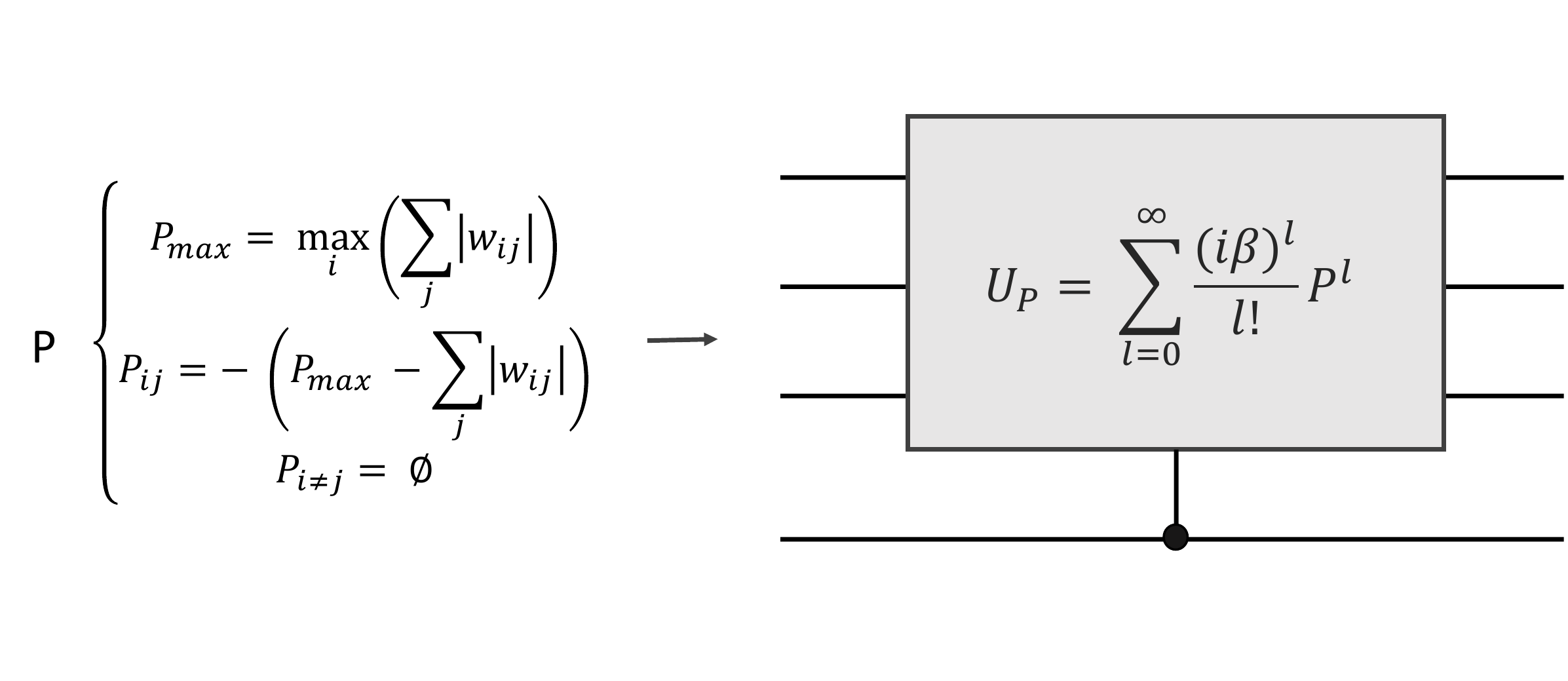} 
        \caption{\textbf{Controlled-unitary population balancing}}
        \label{fig:main-b}
    \end{subfigure}
    \newline
    \vspace{10pt}
    
    \begin{subfigure}{.5\textwidth}
      \centering
        \includegraphics[height=2cm]{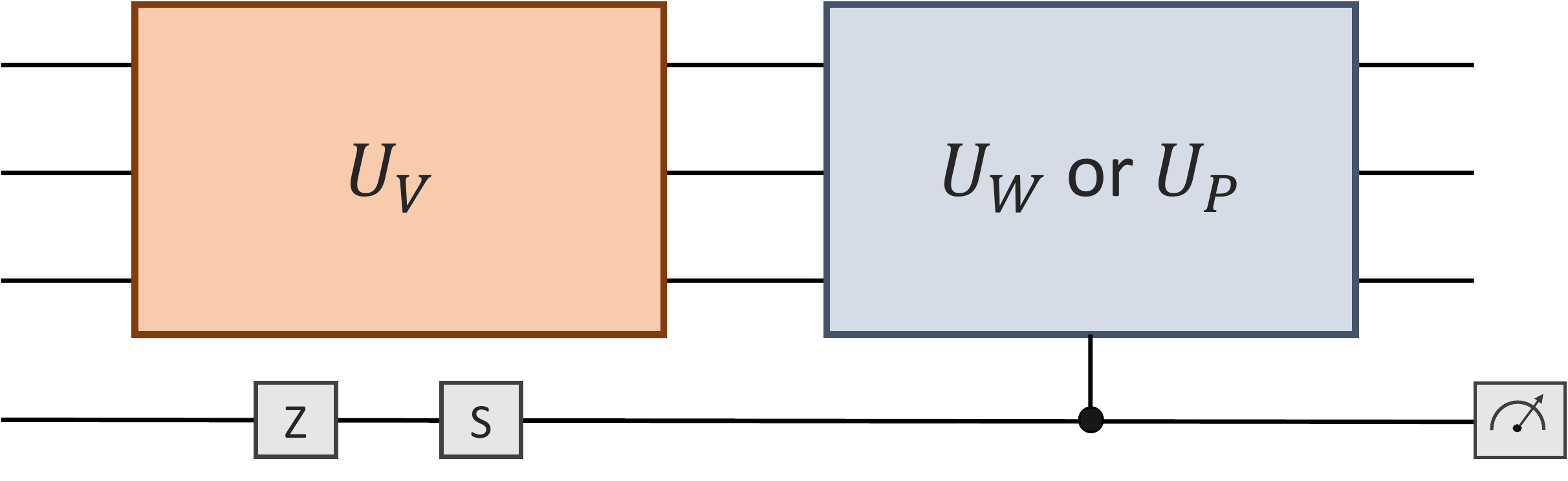}  
        \caption{\textbf{Estimating Hadamard Test loss function terms}}
        \label{fig:main-c}
    \end{subfigure}
    \hfill
    \begin{subfigure}{.5\textwidth}
      \centering
        \includegraphics[height=2cm]{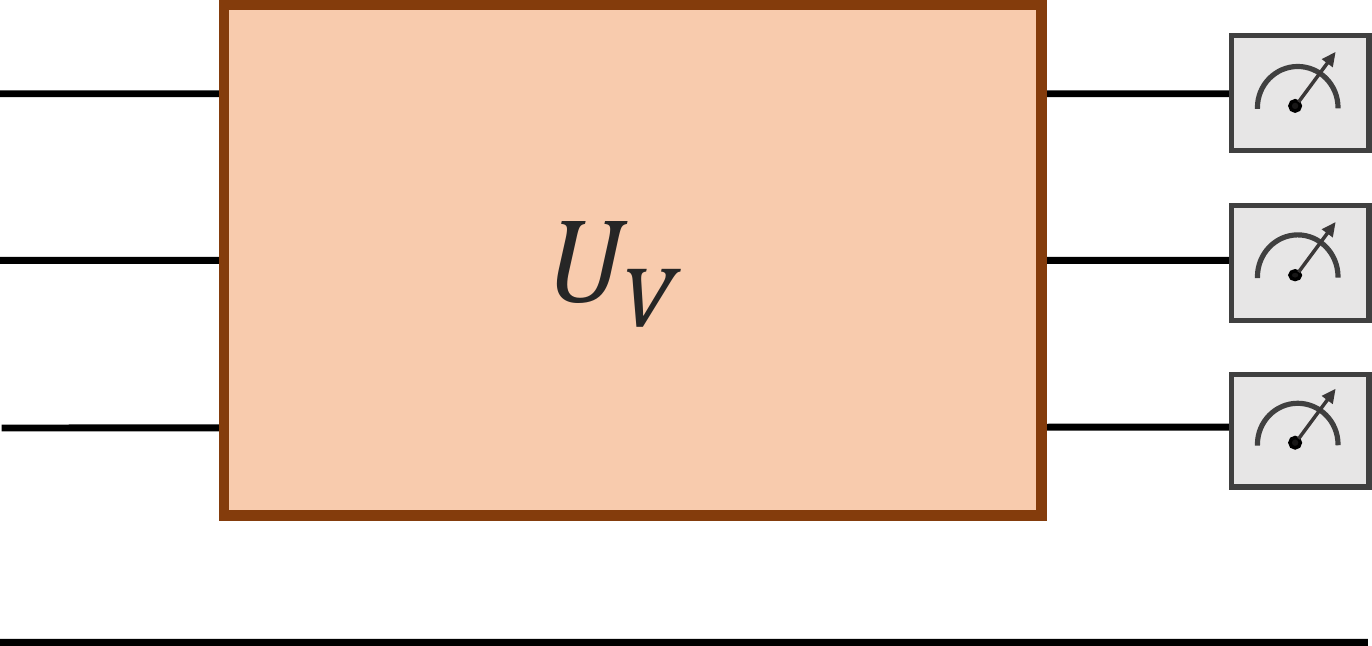}  
        \caption{\textbf{Estimating Pauli string amplitude constraints}}
        \label{fig:main-d}
    \end{subfigure}
    \caption{\textbf{Diagram of HTAAC-QSDP} for the Goemans-Williamson algorithm with $n=3$ non-auxiliary qubits.
    (\ref{fig:main-a}) A classical problem of $N$ variables (here an $N$-vertex MaxCut problem where $N=8$). The weight matrix $W$ is used to generate the unitary $U_W$, which is rotated about the angle $\alpha$ and implemented as a controlled-unitary conditioned on the $n{+}1$th-qubit (auxilary qubit). 
    (\ref{fig:main-b}) The population-balancing unitary $U_P$ is generated by the diagonal matrix $P$, which offsets the asymmetric edge weights on certain vertices in proportion to some constant $\beta$. 
    (\ref{fig:main-c}) The Hadamard Test is used to efficiently evaluate the objective function and population balancing constraints. The $n$-qubit state $|\psi\rangle = U_V|\mathbf{0}\rangle$ is prepared with a variational quantum circuit $U_V$. Although $U_V$ can, in general, be made of any set of parameterized $n$-qubit quantum gates, in this work, we use the circuits described in Sec.\ \ref{sec.simulation}. The $n+1$th (auxilary) qubit is initialized as $(|0\rangle -i |1\rangle)/\sqrt{2}$. Subsequently, the Hadamard Test is carried out: $U_W$ or $U_P$ is implemented as a controlled-unitary conditioned on the auxilary qubit, which is then measured to compute $\langle \sigma_{n+1} \rangle_{W,t} = \text{Im}[\langle \psi|U_W|\psi \rangle]$ or $\langle \sigma_{n+1} \rangle_{P,t} = \text{Im}[\langle \psi|U_P|\psi \rangle]$. 
    (\ref{fig:main-d}) The $M=2^n$ SDP amplitude constraints constraints are approximately enforced with only $m \sim \text{poly}(n)$ Pauli string constraints (Eq.~\ref{eq.constraints}). These are computed by collecting $n$-qubit Pauli-$z$ measurements and using marginal statistics to estimate the $m$ expectation values.}
    \label{fig:main_figure}
\end{figure*}

\section{Introduction}

Semidefinite programming (SDP) is a variety of convex programming wherein the objective function is extremized over the set of symmetric positive semidefinite matrices $\mathbb{S}^+$ \cite{Boyd2004}. Typically, an $N$-variable extremization problem is upgraded to an optimization over $N$ vectors of length $N$, which form the semidefinite matrices of $\mathbb{S}^+$. A versatile technique, SDP can be used to approximately solve a variety of problems, including combinatorial optimization problems (e.g., NP-hard problems, whose computational complexity grows exponentially in problem size) \cite{Goemans1997}, and is heavily used in fields such as operations research, computer hardware design, and networking~\cite{Vandenberghe1999, Li2020}. In many such cases, semidefinite programs (SDPs) are integer programming relaxations, meaning that the original objective function of integer variables is reformed as a function of continuous vector variables \cite{Helmberg2000}, such as the Goemans-Williamson algorithm \cite{Goemans1995}. This allows the SDP to explore a convex approximation of the problem. Although such solutions are only approximate, SDPs are useful because they can be efficiently solved with a variety of techniques. These include interior-point methods, which typically run in polynomial-time in the number of problem variables $N$ and constraints $M$ \cite{Potra2000}. In recent years, more efficient versions of these classical methods have been developed \cite{Jiang2020, Huang2022}.

An additional advantage of optimization with SDPs is that many have performance guarantees in the form of approximation ratios. Approximation ratios are a provable worst-case ratio between the value obtained by an approximation algorithm and the ground truth global optimum \cite{Williamson2011}. In short, SDPs represent an often favorable compromise between computational complexity and solution quality.

\begin{table*}[htbp]
\caption{\label{tab.table0}
Comparison of common quantum methods for classical optimization. The number of potential variables $N$ and constraints $M$ are given in terms of qubits $n$. Whether or not the method provides guarantees on its solutions is discussed, as is its suitability for near-term quantum devices, (i.e., fewer than hundreds of qubits with limited error correction \cite{Moll2018}). Our Hadmard Test objective function and Approximate Amplitude Constraint Quantum SDP (HTAAC-QSDP, Fig.~\ref{fig:main_figure}) ensures SDP approximation ratios, is suitable for near-term variational quantum devices, and provides efficient objective function evaluation (via the Hadamard Test) and constraints (via a second Hadamard Test and $O(n^2)$ Pauli string constraints).
}
\centering
%\begin{ruledtabular}
\begin{tabular*}{1\textwidth}{@{\extracolsep{\fill}} ccccccc}
\toprule
\textbf{Method}&
\textbf{$N$, $M$ Scaling}&
\textbf{Solution Guarantee}&
\textbf{Near-Term Devices}
\\
%\midrule
Quantum Adiabatic \cite{Farhi2000,Albash2018,Ebadi2022} & $n$ & If Infinitely Slow & Sometimes\\
Quantum Annealing \cite{Kadowaki1998,Gibney2017} & $n$ & No & Yes\\
QAOA \cite{Farhi2014} & $n$ & Sometimes & Yes\\
Boson Sampling \cite{Arrazola2021} & $n$ & No & Yes\\
QSDPs \cite{Brandao2019, Apeldoorn2019, Apeldoorn2020, Brandao2017, Brandao2022} & $2^n$ & SDP Approx. Ratios & No\\
Variational QSDPs \cite{Patel2021} & $2^n$ & SDP Approx. Ratios & Yes\\
\textbf{HTAAC-QSDP (Ours)} & $2^n$ & SDP Approx. Ratios & Yes, $ \text{poly}(n)$ exp. vals./epoch
%\bottomrule
\end{tabular*}
%\end{ruledtabular}
\end{table*}

Despite the favorable scaling of classical SDPs, they still become intractable for high-dimensional problems. A variety of quantum SDP algorithms (QSDPs) that sample $n$-qubit Gibbs states to solve SDPs with up to $N=2^n$ variables and $M \sim O(N)$ constraints have been devised \cite{Brandao2019, Apeldoorn2019, Apeldoorn2020, Brandao2017, Brandao2022} (Table~\ref{tab.table0}), as have methods for approximately preparing Gibbs states with near-term variational quantum computers \cite{chowdhury2020variational, Patti2022MCMC, wang2021variational}. The former of these algorithms are based on the Arora-Kale method \cite{Arora2012} and provide up to a quadratic speedup in $N$ and $M$. However, they scale significantly poorer in terms of various algorithm parameters, such as accuracy, and are not suitable for near-term quantum computers. Quantum interior-point methods have also been proposed \cite{Kerenidis2020, Augustino2021}, in close analogy to the leading family of classical techniques.

Variational methods have long played a role in quantum optimization protocols \cite{cerezo2020variational} (Table~\ref{tab.table0}), such as adiabatic computation \cite{Farhi2000,Albash2018,Ebadi2022}, annealing \cite{Kadowaki1998,Gibney2017}, the Quantum Approximate Optimization Algorithm (QAOA) \cite{Farhi2014}, and Boson Sampling \cite{Arrazola2021}. However, only recently have variational QSDPs been proposed \cite{Patel2021,Bharti2021}. Patel et al \cite{Patel2021} addresses the same optimization problems as the quantum Arora-Kale and interior-point based methods, but instead uses variational quantum circuits, which are more feasible in the near-term. Like other SDPs, this method offers specific performance guarantees in the form of approximation ratios \cite{Williamson2011}. While exact methods are efficient for some SDPs, for worst-case problems (e.g., problems with a large number of constraints $M$, such as MaxCut) they may require the estimation of up to $O(2^n)$ observables per training epoch. Although it has been demonstrated that solving NP-hard optimization problems on variational quantum devices does not mitigate their exponential overhead, problems such as MaxCut may still retain APX hardness and are upper bounded by the same approximation ratio of classical methods \cite{Bittel2021}.
Likewise, while parameterized quantum circuits can form Haar random $2$-designs that result in barren plateaus \cite{McClean2018}, numerous methods of avoiding, mitigating, or and perturbing these systems to effectuate a more trainable space have been developed \cite{Marrero2021,Patti2021,Pesah2021}.

\begin{figure*}[t!]
\centering
\includegraphics[width=2\columnwidth]{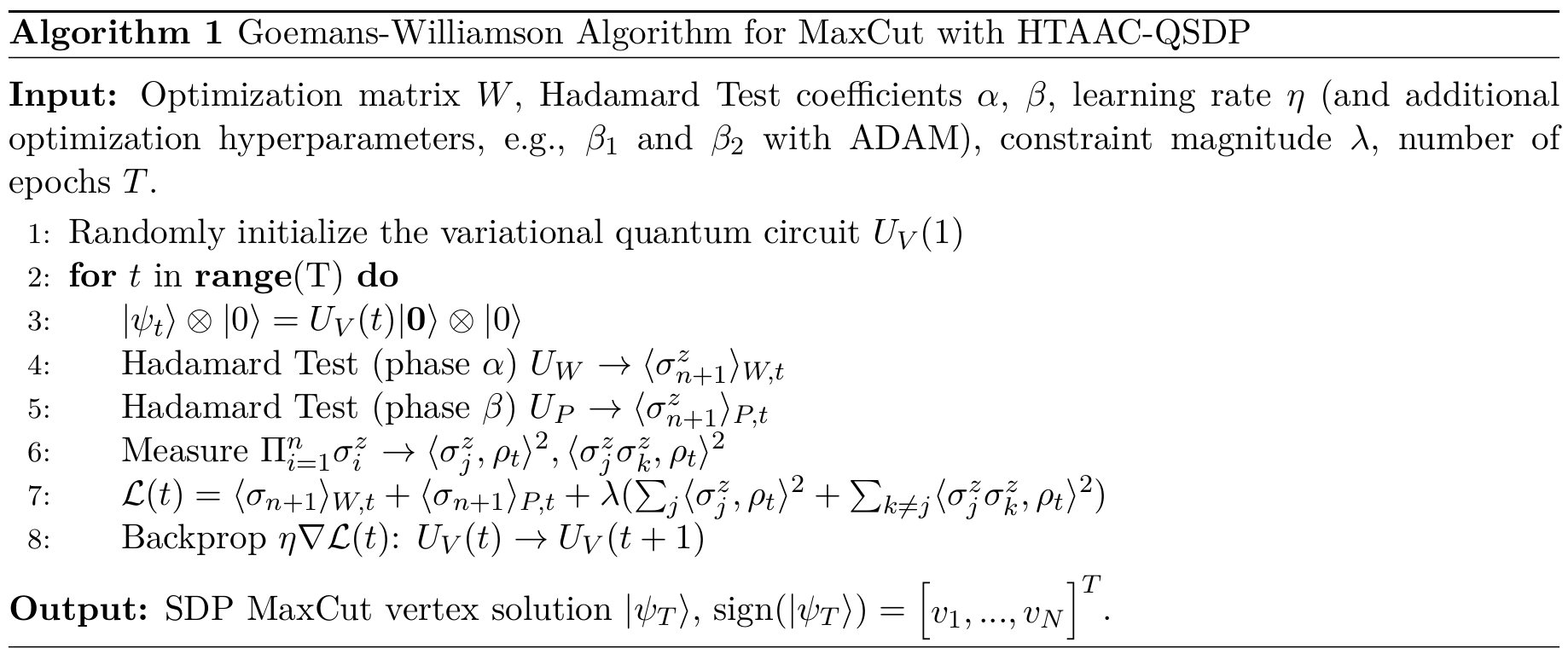}
\end{figure*}

\textbf{Our Approach:} We propose a new variational quantum algorithm for approximately solving QSDPs that uses Hadamard Test objective functions and Approximate Amplitude Constraints (HTAAC-QSDP, Fig.~\ref{fig:main_figure}). 
\begin{theorem}
    HTAAC-QSDP for the Goemans-Williamson algorithm uses $n{+}1$ qubits, a constant number of quantum measurements, and $O(\text{poly}(n))$ classical calculations to approximate SDPs with up to $N=2^n$ variables.
\end{theorem}
The details of the HTAAC-QSDP implementation must be engineered to each SDP and, in this work, we focus on the Goemans-Williamson algorithm \cite{Goemans1995}, with particular emphasis on its application to MaxCut. In some cases, our method is nearly an exponential reduction in required expectation values, e.g., for high-constraint problems such as MaxCut. As described in Sec.~\ref{sec.algorithm}, we achieve this, in part, through a unitary objective function encoding with the Hadamard Test \cite{Aharonov2009} (Fig.~\ref{fig:main-a}), which allows for the extremization of the entire $N$-dimensional objective by estimating only a single quantum expectation value (Fig.~\ref{fig:main-c}). Our quantum Goemans-Williamson algorimth for MaxCut requires $M = N \leq 2^n$ amplitude constraints, which we effectively enforce with only 1) a constant number of quantum measurements from a second Hadamard Test (Fig.~\ref{fig:main-b}) and 2) the estimation of a polynomial number of properly selected, commuting Pauli strings (Fig.~\ref{fig:main-d}).

In Sec.~\ref{sec.results}, we demonstrate the success of the HTAAC-QSDP Goemans-Williamson algorithm (Algorithm 1) by approximating MaxCut \cite{Commander2009} for large-scale graphs from the well-studied GSet graph library \cite{Benson1999} (Fig.~\ref{fig.graphs}). In addition to satisfying the $0.878$ MaxCut approximation ratio \cite{Goemans1995}, HTAAC-QSDP achieves cut values that are commensurate with the leading gradient-based classical SDP solvers \cite{Choi2000}, implying that we reach optima that are very near the global optimum of these SDP objective functions.

Finally, in Sec.~\ref{sec.scale} we discuss the feasibility of the Hadamard encoding. For general SDPs, we establish an upper bound (Theorem~\ref{theorem.general_bound}) on the phase $\alpha$ of our Hadamard encoding, such that our technique is a high-quality approximation of the original SDP. The purpose of this upper bound is to demonstrate that tractably large values of the unitary phase $\alpha$ are permissible (i.e., $\alpha$ need not become arbitrarily small) for encoding a wide variety of large-scale problems. We discuss the known difficulty, and thus usefulness, of graph optimization problems under these conditions. Specifically:

\begin{theorem}
Our approximate Hadamard Test objective function $U_W \sim i \alpha W$ (Sec.~\ref{sec.hadamard}) holds for graphs with randomly distributed edges if

$$\alpha^2 \lesssim \frac{N^4}{e^3} = \frac{N}{\xi^3},
    \label{theorem.general_bound}$$

\noindent where $e$ is the number of non-zero edge weights and $\xi$ is the average number of edges per vertex.
\end{theorem}

We can view the criteria of Theorem~\ref{theorem.general_bound} in two ways: for SDPs with arbitrarily many variables $N$, the size of $\alpha$ can be kept reasonably large while the Hadamard Test objective function (see Sec.~\ref{sec.hadamard}) remains valid as long as 1) $N$ does not grow slower than the total number of edges $e$, or 2) $N$ does not grow slower than the the cube of $\xi$. Both of these conditions hold for graphs that are not too dense, meaning that they are widely satisfiable. For instance, the majority of interesting and demonstrably difficult graphs for MaxCut are relatively sparse \cite{Commander2009,Williamson2011,Benson1999,Wiegele2007,DIMACS}. We note that for graphs where edge-density is unevenly distributed, Theorem~\ref{theorem.general_bound} should hold for the densest region of the graph, i.e., $\xi$ should be the average number of edges per vertex for the most highly connected vertices.

\section{Efficient Quantum Semidefinite Programs}

\label{sec.algorithm}

The standard form of an $N$-variable, $M$-constraint SDP is \cite{Boyd2004, Goemans1997}

\begin{equation}
\begin{split}
    &\text{minimize}_{X \in \mathbb{S}^+} \hspace{0.2cm} \langle W, X \rangle \\
    &\text{subject to  } \langle A_\mu, X \rangle = b_\mu, \hspace{0.2cm} \forall \mu \leq M \\
    & X \succeq 0,
\end{split}
\label{eq.psd}
\end{equation}

\noindent where $W$ is an $N \times N$ symmetric matrix that encodes the optimization problem and $A_\mu$ ($b_\mu$) are $N \times N$ symmetric matrices (scalars) that encode the problem constraints. $\langle A, B \rangle$ denotes the trace inner product

\begin{equation}
\langle A, B \rangle = \text{Tr}\left[A^T B \right] = \sum_{i,j}^N A_{ij} B_{ij}.
\end{equation}

In this section, we detail a method of efficient optimization of the above SDP objective and constraints using quantum methods (Fig.~\ref{fig:main_figure}), specifically by implementing Hadamard Tests and imposing a polynomial number of Pauli constraints. We provide a concrete example in the form of the Goemans-Williamson \cite{Goemans1995} algorithm for MaxCut \cite{Commander2009}, as summarized in Algorithm 1.

\subsection{The Hadamard Test as a Unitary Objective}
\label{sec.hadamard}

In quantum analogy to the objective function of Eq.\ \ref{eq.psd}, we wish to minimize $\langle W, X \rangle$ over the $n$-qubit density matrices $\rho$, which are positive semidefinite by definition. We generate these quantum ensembles $\rho$ from an initial density matrix $\rho_0$ such that $\rho = U_V \rho_0 U_V^\dagger$, where $U_V$ is a variational quantum circuit. This yields the quantum objective function

\begin{equation}
    \text{minimize  } \langle W, \rho \rangle = \text{Tr}\left[ W \rho \right].
    \label{eq.objective}
\end{equation}

\noindent Throughout most of this work, we consider pure states such that $\rho_0 = |0\rangle \langle 0|$ and $|\psi \rangle = U_V |0 \rangle$, although we detail the case of mixed quantum states in Sec.\ \ref{sec.mixed_quantum_states}. In the case of pure states, Eq.\ \ref{eq.objective} yields

\begin{equation}
    \text{minimize  } \langle \psi | W | \psi \rangle.
\end{equation}

The Hadamard Test (Fig.~\ref{fig:main-c}) is a quantum computing subroutine for arbitrary $n$-qubit states $|\psi \rangle$ and $n$-qubit unitaries $U$ \cite{Aharonov2009}. It allows the real or imaginary component of the $2^n$-state inner product $\langle \psi | U | \psi \rangle$ to be obtained by estimating only a single expectation value $\langle \sigma_{n+1}^z \rangle$, which is the $z$-axis Pauli spin on the $n{+}1$th (auxiliary) qubit. For example, to obtain the imaginary component of $\langle \psi | U | \psi \rangle$, we prepare the quantum state $|\psi\rangle \otimes \frac{1}{\sqrt{2}}(|0\rangle -i|1\rangle)$ and apply a controlled-$U$ from the $n{+}1$th qubit to $|\psi\rangle$, followed by a Hadamard gate on the $n{+}1$th qubit. This produces the state

\begin{equation}
    \frac{1}{2}\left[(I - iU) |\psi\rangle \otimes |0\rangle  + (I + iU) |\psi\rangle \otimes |1\rangle \right]
\end{equation}

\noindent upon which projective measurement yields

\begin{equation}
    \langle \sigma_{n+1}^z \rangle = \text{Im}\left[\langle \psi | U | \psi \rangle \right].
\end{equation}

Rather than estimate the $N \leq 2^n$ expectation values required to characterize $\rho$ and optimize the loss function of Eq.~\ref{eq.objective}, our method encodes the $N$-dimensional objective matrix $W$ as the imaginary part of an $n$-qubit unitary $U_W=\exp(i \alpha W)$ (Fig.~\ref{fig:main-a}). Here, the phase $\alpha$ is a constant scalar. $U_W$ is then conditioned on the $n{+}1$th (or auxilary) qubit as a controlled-unitary. We then use the Hadamard Test to calculate the objective term in the loss function

\begin{equation}
\langle \sigma_{n+1}^z \rangle_W = \text{Im}\left[\langle \psi | U_W | \psi \rangle \right] = \text{Im}\left[ \langle U_W, \rho \rangle \right].
\label{eq.unitary_objective}
\end{equation}

The intuition for this objective function is that, for sufficiently small $\alpha$, $\text{Im}[U_W] \approx \alpha W$. By restricting ourselves to quantum circuits with real-valued output states, we render the single expectation value $\langle \sigma_{n+1}^z \rangle_W$ proportional to the objective function of Eq.\ \ref{eq.objective}, which requires $N$ expectation values to estimate. In Sec.~\ref{sec.scale}, we analytically prove that, for many optimization problems, $\alpha$ has a practical upper bound such that $\text{Im}[U_W] \approx \alpha W$ with a reasonably large $\alpha$, even for arbitrarily large $W$.

\begin{figure}[t!]
\centering
\includegraphics[width=1\columnwidth]{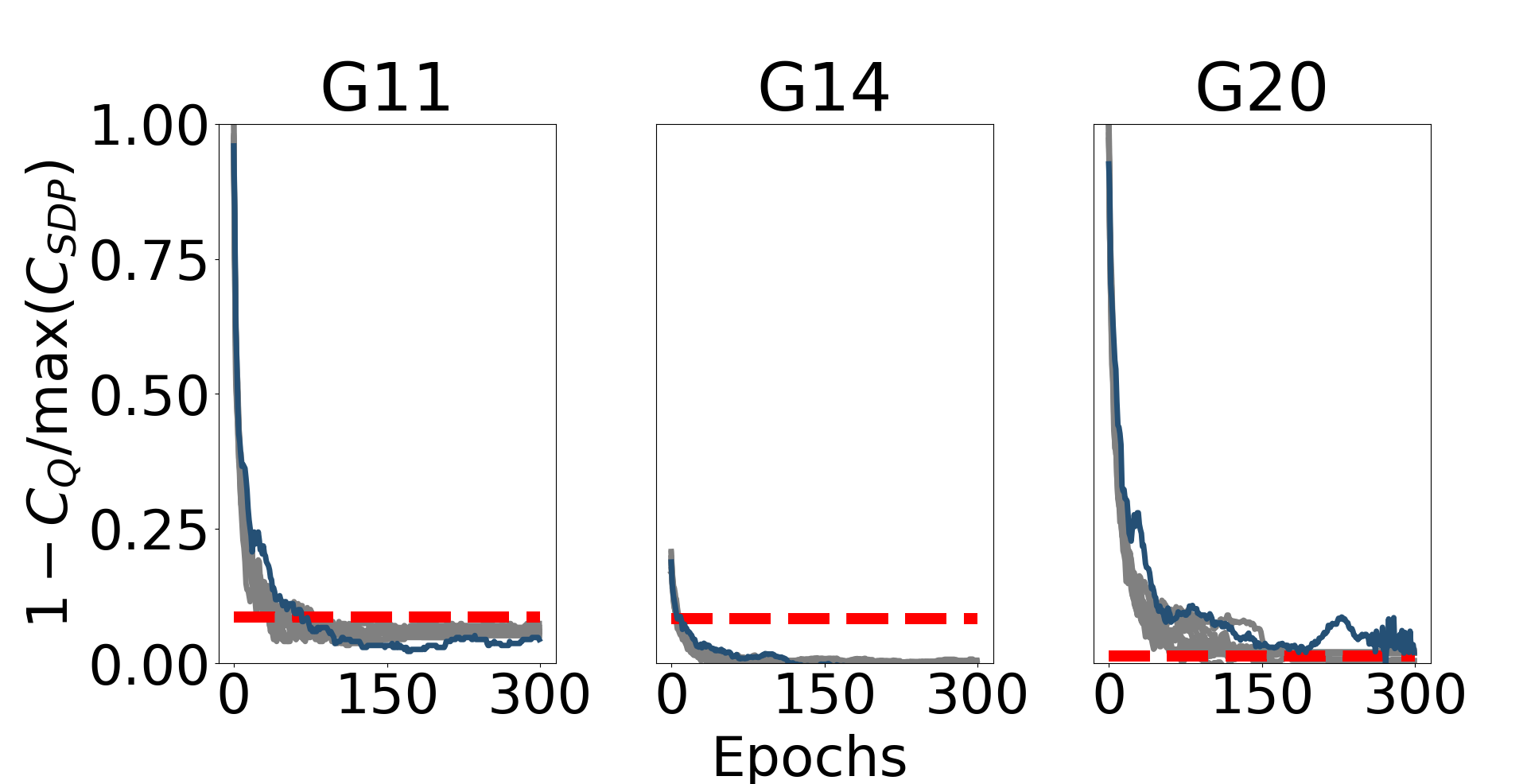}
\caption{The cut values $C_Q$ obtained by the HTAAC-QSDP Goemans-Williamson algorithm with order-$k \leq 2$ Pauli constraints compared to $\text{max}(C_\text{SDP})$, the best results of classical gradient-based SDPs (specifically, interior points methods) \cite{Choi2000}. Our performance on the skewed binary and integer graphs narrowly exceeds that of the classical method ($\text{max}(C_\text{SDP})<C_Q$), while the classical method narrowly outperforms our quantum method for the toroid graphs ($\text{max}(C_\text{SDP})>C_Q$). Overall, the performance of our HTAAC-QSDP implementation and its classical counterpart are commensurate. HTAAC-QSDP exceeds the $C_Q/C_\text{MAX}>0.878$ MaxCut approximation ratio (red dashed line) for all graphs, where $C_\text{MAX}$ is the true MaxCut of the graph. In this work, we assume $C_\text{MAX}$ as the largest-known cuts of the GSet graphs, which were obtained from intensive and repeated heuristic searches \cite{Toshiba}. The quantum mixed state implementation detailed in Sec.\ \ref{sec.mixed_quantum_states} is depicted in dark blue. It furnishes a higher rank solution and has marginally improved performance.}
\label{fig.cut_vs_epochs}
\end{figure}

\subsection{Quantum Goemans-Williamson Algorithm}

We now illustrate how $\text{Im}[U_W]$ can be a close approximation of $\alpha W$, including for optimization problems with an arbitrarily large number of variables $N$. For concreteness, we select the NP-complete problem MaxCut \cite{Commander2009}, and specifically focus on the corresponding NP-hard optimization problem~\cite{Karp1972}. This problem is of particular interest due to its favorable $0.878$-approximation ratio with semidefinite programming techniques, notably the Goemans-Williamson algorithm \cite{Goemans1995}, for which we now derive an efficient quantum implementation. The Goemans-Williamson algorithm is also applicable to to numerous other optimization problems, such as MaxSat and Max Directed Cut \cite{Goemans1995}.

For a MaxCut problem with $N$ vertices $v_i$, $v_j$, let $W$ be the matrix that encodes the up to $N(N-1)/2$ non-zero edge weights in its entries $W_{ij}$. As the vertices lack self-interaction, $W_{ii}=0$. The optimization problem is then defined as

\begin{equation}
\begin{split}
&\text{maximize  } \sum_{j < i} W_{ij} \frac{1-v_i v_j}{2} \\
&\text{subject to  } v_i = \pm 1,
\label{eq.maxcut_combinatorial}
\end{split}
\end{equation}

\noindent which can be mapped to the classical SDP with $M=N$ constraints

\begin{equation}
\begin{split}
    &\text{minimize}_{X \in \mathbb{S}^+} \hspace{0.2cm} \langle W, X \rangle \\
    &\text{subject to} \hspace{0.2cm} X_{ii} = 1, \hspace{0.2cm} \forall i \leq N.
    \label{eq.maxcut}
\end{split}
\end{equation}

\begin{table*}[htbp]
\caption{\label{tab.table1}
MaxCut statistics for all $800$-vertex graphs studied by the leading gradient-based classical SDP (interior points) method \cite{Choi2000}. The highest known MaxCut values ($C_\text{MAX}$, found by intensive heursitics \cite{Toshiba}) are greater the highest results obtained by the classical method ($\text{max}(C_\text{SDP})$), but the approximation ratio $\text{max}(C_\text{SDP})/C_\text{MAX} > 0.878$ is satisfied. The largest cut values of our HTAAC-QSDP Goemans-Williamson algorithm ($\text{max}(C_Q)$) are comparable with $\text{max}(C_\text{SDP})$, as are the average results ($\text{mean}(C_Q)$).
}
\centering
%\begin{ruledtabular}
\begin{tabular*}{1\textwidth}{@{\extracolsep{\fill}} ccccccc}
\toprule
\textbf{Graph}&
\textbf{$C_\text{MAX}$}&
\textbf{$\text{max}(C_\text{SDP})$}&
\textbf{$\text{max}(C_Q) / \text{max}(C_\text{SDP})$}&
\textbf{$\text{mean}(C_Q) / \text{max}(C_\text{SDP})$}
\\
%\midrule
G11 & $564$ & $542$ & $0.967$ & $0.940$\\
G12 & $556$ & $540$ & $0.982$ & $0.953$\\
G13 & $582$ & $564$ & $0.972$ & $0.933$\\
G14 & $3064$ & $2922$ & $1.011$ & $1.000$\\
G15 & $3050$ & $2938$ & $1.009$ & $0.996$\\
G20 & $941$ & $838$ & $1.007$ & $0.983$\\
G21 & $931$ & $841$ & $1.001$ & $0.978$
%\bottomrule
\end{tabular*}
%\end{ruledtabular}
\end{table*}

As described by Eq.~\ref{eq.objective}, we can transform the optimization portion of Eq.~\ref{eq.maxcut} by substituting the classical positive semidefinite matrix $X$ for the quantum density operator $\rho$. The solution to the SDP is then stored in $|\psi\rangle$, i.e., $v_i = \text{sign}(\psi_i)$ (for more details, see Sec.~\ref{sec.solution}). As detailed in Sec.~\ref{sec.hadamard}, the evaluation of this objective function can be optimized by estimating a single expectation value with the Hadamard Test. Likewise, we now introduce a quantum alternative to the constraint $X_{ii}=1$ from Eq.\ \ref{eq.maxcut}. First, note that due to the orthonormality of quantum states, the exact quantum equivalent of Eq.\ \ref{eq.maxcut} is

\begin{equation}
\rho_{ii}=1/2^n \leq N^{-1}.
\label{eq.quantum_norm}
\end{equation}

\noindent This rescaling changes neither the effectiveness nor the guarantees of the semidefinite program, because the salient feature of the constraint is that all of the quantum states have the same amplitude magnitude $|\psi_i|$, such that all of the vertices are of equal magnitude and none are disproportionately favored. The solutions $\rho$ and $X$ differ only by a constant factor, such that $\rho=X/2^n$. This yields the quantum MaxCut SDP

\begin{equation}
\begin{split}
    &\text{minimize} \hspace{0.2cm} \langle W, \rho \rangle \\
    &\text{subject to} \hspace{0.2cm} \rho_{ii} = 2^{-n}, \hspace{0.2cm} \forall i \leq N.
    \label{eq.quantum_maxcut}
\end{split}
\end{equation}

As graph weights are real-valued and symmetric (i.e., $W_{ij}=W_{ji}$), $W$ is Hermitian. We can thus use it as the generator of $U_W$ such that (Fig.~\ref{fig:main-a})

\begin{equation}
\begin{split}
    &U_W = \exp(i \alpha W) = \sum_l \frac{(i\alpha)^l}{l!} W^l \\
    &= I + \frac{i\alpha}{1!} W - \frac{\alpha^2}{2!} W^2 - \frac{i\alpha^3}{3!} W^3 + \mathcal{O}(W^4).
    \label{eq.unitary}
\end{split}
\end{equation}

\noindent As $W$ is real, the odd powers of $l$ in Eq.\ \ref{eq.unitary} are the imaginary components. The condition that $\text{Im}[U_W] \propto W$ is upheld \textit{iff}, for the vast majority of variables $i,j$, $\alpha W_{ij} \gg \frac{\alpha^3}{6}(W^3)_{ij}$. Note that, when this condition holds, $\langle \sigma^z_{n+1} \rangle_W$ approximates $\langle W \rangle$ with only vanishing error and a rescaling by $\alpha$. In Sec.~\ref{sec.scale}, we prove Theorem~\ref{theorem.general_bound}, demonstrating that this condition is achievable with a tractable $\alpha$ (i.e., $\alpha$ larger than some fixed finite value that is constant in problem size $N$) for a wide variety of graphs.

Next, we note that enforcing the $M=N \leq 2^n$ amplitude constraints $\rho_{ii}=2^{-n}, \hspace{0.1cm} i \leq N$ would require the estimation of all $z$-axis Pauli strings of order $k \leq n$ (all Pauli strings with $k \leq n$ Pauli-$z$ operators) of the state $|\psi \rangle$. This would be a total of $N-1$ expectation values. As an alternative to this large overhead, HTAAC-QSDP proposes the use of \textit{Approximate Amplitude Constraints} (Fig.~\ref{fig:main-d}). For example, consider the set of $m = n(n-1)/2 + n \sim n^2/2$ Pauli strings of length $k \leq 2$

\begin{equation}
\begin{split}
    &\langle \sigma_a^z, \rho \rangle = 0, \hspace{0.2cm} \forall a \leq n \\
    &\langle \sigma_a^z \sigma_b^z, \rho \rangle = 0, \hspace{0.2cm} \forall b\neq a, \hspace{0.2cm} a,b \leq n,
    \label{eq.constraints}
\end{split}
\end{equation}

\noindent where $\sigma_a^z$ are the Pauli $z$-operators on the $a$th qubit. We can use these equalities as partial constraints for the $n$-qubit output state $|\psi\rangle$. This set of $m \sim n^2/2$ constraints approximates the same restrictions as the set of $M=N$ constraints of Eq.~\ref{eq.quantum_maxcut} by limiting quantum correlations, as these indicate subsets of states with unequal populations. That is, each such $z$-axis Pauli string is the difference of the total populations of two equal partitions of state space. To gain intuition, let us consider a two-qubit state $|\psi \rangle = [\psi_{00}, \psi_{01}, \psi_{10}, \psi_{11}]^T$ and the $k=1$ Pauli string $\sigma_0^z \otimes I$. Using the constraints for Eq.\ \ref{eq.constraints}, we enforce the equality $|\psi_{00}|^2 + |\psi_{01}|^2 = |\psi_{10}|^2 + |\psi_{11}|^2$ which, while not fully enforcing the constraints of Eq.\ \ref{eq.quantum_maxcut} (e.g., not enforcing that \textit{all} states have equal populations), does enforce equal populations among a subset of states. This results in a lighter-weight and more flexible set of constraints. As an example, $\sigma_0^z \otimes I$ promotes all state components to be populated by disallowing states where the first qubit is in a computational basis state, such as $|00\rangle$ or $\frac{1}{\sqrt{2}}(|00\rangle + |01\rangle)$. Likewise, the $k=2$, $z$-axis Pauli string $\sigma_0^z \otimes \sigma_1^z$ constraint produces the equality $|\psi_{00}|^2 + |\psi_{11}|^2 = |\psi_{01}|^2 + |\psi_{10}|^2$, which would, for example, disallow the Bell State $(|00\rangle + |11\rangle)/\sqrt{2}$. We highlight that the Pauli string constraints of Eq.~\ref{eq.constraints} are commuting, such that they can be estimated as $m$ different marginal distributions from a single set of $n$-qubit $z$-axis measurements.

We again emphasize that these $k \leq 2$ constraints only approximately enforce the SDP constraint $\rho_{ii}=2^{-n}$. Fully satisfying this constraint would require restricting Pauli-$z$ correlations between \textit{any} subset of the $n$ qubits, such that no states of unequal amplitude magnitudes are permitted. Eq.~\ref{eq.quantum_norm} can be fully satisfied if we constrain $| \psi \rangle$ with all of the Pauli strings of length $k \leq n$. However, as there exist $n$ choose $k$ $z$-axis Pauli strings of order $k$, this requires estimating $\sum_{k=1}^n \left(\begin{smallmatrix}n \\ k\end{smallmatrix}\right) = 2^n -1$ different expectation values and greatly decreases the efficiency of the algorithm. Sec.~\ref{sec.results} details that, in practice for $800$-vertex graphs, competitive results are obtained using only $k \leq 2$ constraint terms (Fig.~\ref{fig.cut_vs_epochs} and Table~\ref{tab.table1}), and optimization performance is largely saturated with terms $k \leq 4$ (Fig.~\ref{fig.sigmarhoii_vs_epochs} and Table~\ref{tab.table2}).

\begin{figure}[t!]
\centering
\includegraphics[width=1\columnwidth]{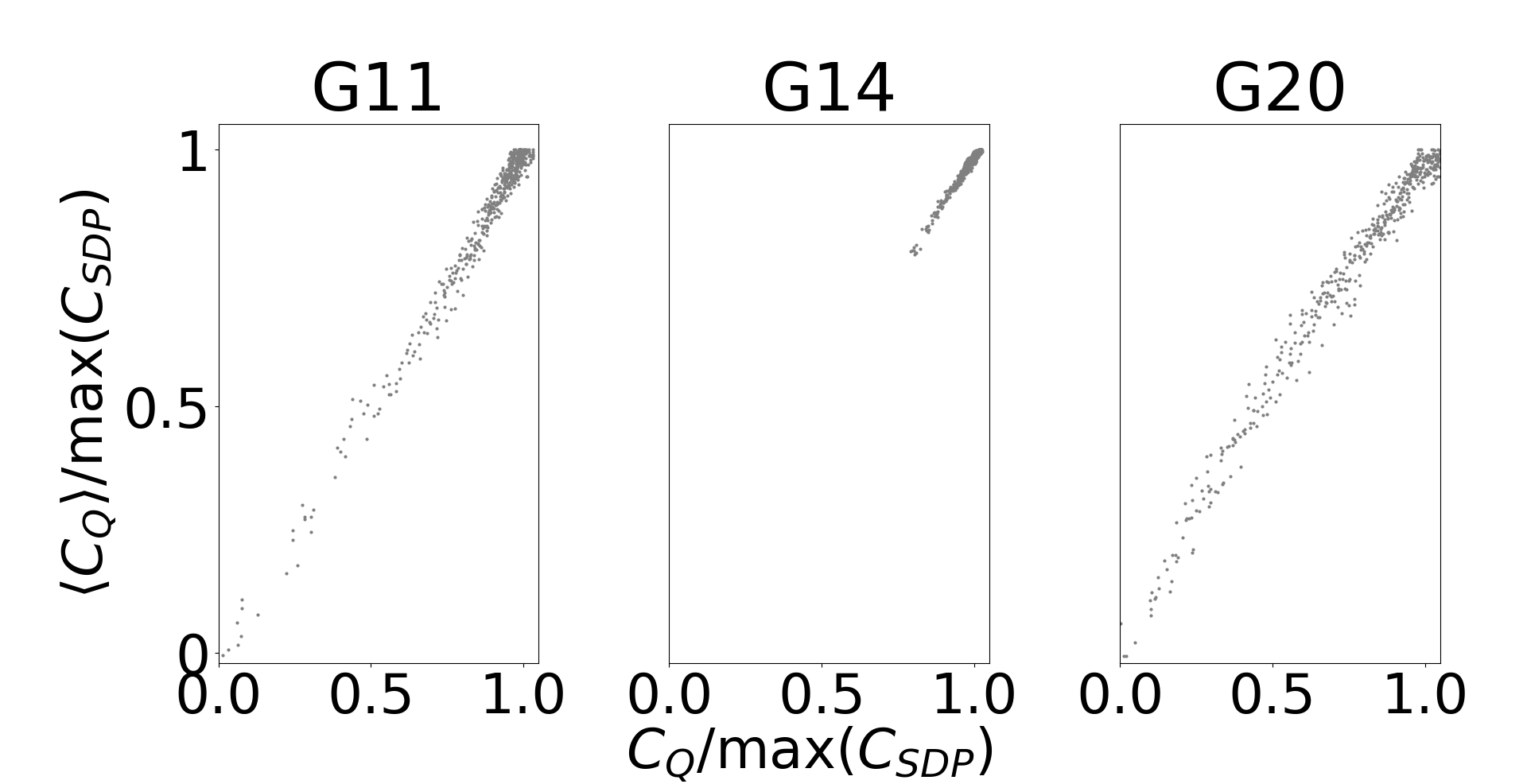}
\caption{The cut value estimated by quantum observables $\langle C_Q \rangle$ obtained by HTAAC-QSDP vs the true SDP rounded cut value $C_Q$ for the G11 (toroid), G14 (binary), and G20 (integer) graphs. As with classical SDP methods, low loss function values are correlated with high cut values. The strong correlation between the observable estimated cut and the rounded true cut value not only illustrates the convergence of the HTAAC-QSDP Goemans-Williamson algorithm despite its approximate nature, it also demonstrates its ability to extract cut values of many variables from few measurements.}
\label{fig.cut_vs_loss}
\end{figure}

In order to explicitly see how the $m\sim n^2/2$ constraints of Eq.\ \ref{eq.constraints} largely enforce the constraint $\rho_{ii}=1/2^n$, let us take the example of a three-qubit state ($n=3$), which can encode up to eight vertices ($N=2^n=8$) using HTAAC-QSDP. Any real-valued $n=3$ state can be written generically as

$$|\psi \rangle = \sum_{r, s, p=0}^1 \psi_{rsp} |\psi_{rsp} \rangle$$

%$$|\psi \rangle = \psi_{000} |00 \rangle + \beta |01\rangle + \gamma |10\rangle + \delta |11\rangle$$,

\noindent and its constraints from Eq.~\ref{eq.constraints} are

\begin{equation}
\begin{split}
    &\langle \sigma_1^z, \rho \rangle = \sum_{r,s}|\psi_{0rs}|^2 - \sum_{r,s}|\psi_{1rs}|^2 = 0 \\
    &\langle \sigma_2^z, \rho \rangle = \sum_{r,s}|\psi_{r0s}|^2 - \sum_{r,s}|\psi_{r1s}|^2 = 0 \\
    &\langle \sigma_3^z, \rho \rangle = \sum_{r,s}|\psi_{rs0}|^2 - \sum_{r,s}|\psi_{rs1}|^2 = 0 \\
    &\langle \sigma_1^z \sigma_2^z, \rho \rangle = \sum_p ( \sum_{s = r}|\psi_{rsp}|^2 - \sum_{s \neq r}|\psi_{rsp}|^2 ) = 0 \\
    &\langle \sigma_1^z \sigma_3^z, \rho \rangle = \sum_p ( \sum_{s = r}|\psi_{rps}|^2 - \sum_{s \neq r}|\psi_{rps}|^2 ) = 0 \\
    &\langle \sigma_1^z \sigma_2^z, \rho \rangle = \sum_p ( \sum_{s = r}|\psi_{prs}|^2 - \sum_{s \neq r}|\psi_{prs}|^2 ) = 0.
\end{split}
\end{equation}

\noindent Combined with the normalization constraint $\langle \psi|\psi \rangle = 1$, the above system of equations \textit{nearly} guarantees that Eq.~\ref{eq.quantum_norm} is fulfilled. However, it still permits a small subset of states that do not satisfy Eq.~\ref{eq.quantum_norm} due to three-qubit correlations, e.g.,

$$|\psi^* \rangle = \frac{1}{2}\begin{bmatrix} \pm 1, 0, 0, \pm 1, 0, \pm 1, \pm 1, 0 \end{bmatrix}^T.$$

\noindent States with higher-order correlations such as $|\psi^*\rangle$, which neither satisfy Eq.~\ref{eq.quantum_norm} nor are disallowed by Eq.~\ref{eq.constraints}, can be avoided by adding higher-order Pauli string constraints. For the above $n=3$ example, we would add the $k=n=3$ constraint $\langle \sigma_1^z \sigma_2^z \sigma_3^z, \rho \rangle = 0$, which would disallow $|\psi^*\rangle$ as $\langle \sigma_1^z \sigma_2^z \sigma_3^z, \rho^* \rangle = 1$.

%$$|\alpha|^2 = |\beta|^2 = |\gamma|^2 = |\delta|^2 = 1/4 = 1/2^n.$$

Eq.~\ref{eq.quantum_norm} can also be systematically undermined by the unequal distribution of graph edges among the quantum states. For instance, the asymmetrically distributed edge-weights in skewed graphs (Fig.~\ref{fig.graphs} left and Sec.~\ref{sec.results}). With such graphs, the minimization of the loss function can lead to outsized state populations for quantum states that encode high-degree (high edge-weight) vertices. Moreover, as the number of classical variables will not generally be a power of two, there will often be quantum states that are not encoded with a classical variable. For example and as detailed in Sec.~\ref{sec.results}, we use $n=10$ qubits ($2^n=1024$ states) to optimize the $800$-vertex ($N=800$) GSet graphs, such that the states $801$ to $1024$ are absent from the objective function. In such cases, the minimization of the loss function can lead to outsized state populations of quantum states that are present in the optimization function. In principle, these imbalances can be addressed by increasing the magnitude of the Pauli string amplitude constraints, but this is known to cause poor objective function convergence \cite{Bertsekas2014}.

To redress this systematic skew, we add a population-balancing unitary $U_P$ (Fig.~\ref{fig:main-b}), which is implemented on $|\psi \rangle$ via a second Hadamard Test (Sec.~\ref{sec.hadamard}, Fig.~\ref{fig:main-c}) and adds the loss function term $\langle \sigma_{n+1}\rangle_P$. Specifically, $U_P = \exp(i\beta P)$ where $P$ is some diagonal operator of edge weights $P_{ii} = - (P_{max}-\sum_j |\omega_{ij}|)$, where $\beta$ is an adjustable hyperparameter and $P_{max} = \text{max}_i(\sum_j|\omega_{ij}|)$ is the maximum magnitude of edge weights for any given vertex. $U_P$ works to balance the state populations by premiating the occupation of states that are lesser represented by or absent from the objective function $\langle \sigma_{n+1}\rangle_W$.

\begin{figure}[t!]
\centering
\includegraphics[width=1\columnwidth]{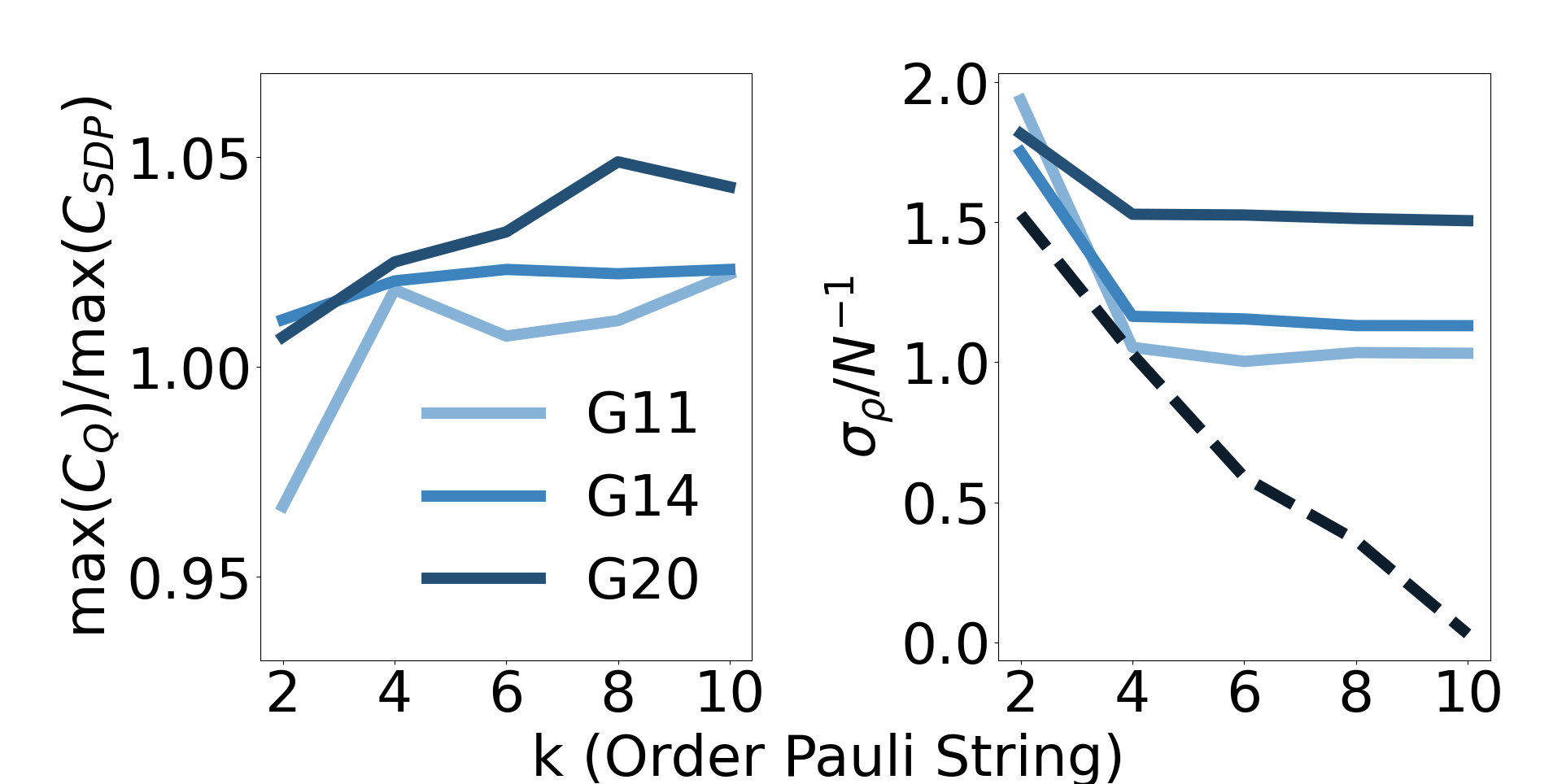}
\caption{The effect of including higher-order HTAAC-QSDP Pauli string amplitude constraints in MaxCut optimization on the G11 (toroid), G14 (binary), and G20 (integer) graphs \cite{Benson1999}. (Left) the performance of HTAAC-QSDP increases as higher-order Pauli strings are used to constrain state amplitudes. The algorithm's performance saturates with $k \approx 4$, indicating that the benefits saturate with less than a polynomial number of Pauli string constraints ($k\leq n$). As illustrated by this work (e.g., Fig.~\ref{fig.cut_vs_epochs} and Table~\ref{tab.table1}), $k=2$ ($m \approx n^2/2$) is often sufficient for competitive SDP optimization. (Right, solid lines) the variance of state magnitude $\sigma_\rho = \text{var}(\rho_{ii}) = \text{var}(|\psi_i|^2)$ vs the order $k$ of Pauli strings constraints. As $k$ increases, $\sigma_{\rho}$ decreases considerably, although this effect is largely saturated by $k \approx 4$. (Right, dashed line) in the absence of competing dynamics (i.e., $\langle \sigma_{n+1}\rangle_W$ and $\langle \sigma_{n+1}\rangle_P$), the Pauli string constraints are fully enforced such that $|\psi_i| \rightarrow N^{-1/2}$ ($\sigma_\rho \rightarrow 0$) as $k\rightarrow n$.}
\label{fig.sigmarhoii_vs_epochs}
\end{figure}

Combining both the efficient Hadamard Test objective function and the Approximate Amplitude Constraints, we can use simple gradient descent-based penalty methods \cite{Bertsekas2014} to find the solution. Specifically, we minimize the HTAAC-QSDP loss function for the Goemans-Williamson algorithm

\begin{equation}
\begin{split}
    &\mathcal{L}(t) = \langle \sigma_{n+1}\rangle_{W,t} + \langle \sigma_{n+1}\rangle_{P,t} \\
    &+ \lambda \left[\sum_j \langle \sigma_j^z, \rho_t \rangle^2 + \sum_{k\neq j} \langle \sigma_j^z \sigma_k^z, \rho_t \rangle^2 \right]
    \label{eq.loss}
\end{split}
\end{equation}

\noindent at each time step $t$ by preparing a quantum state $\rho_t$ on a variational quantum computer. The scalar $\lambda$ is the penalty hyperparameter. While for simplicity we have chosen a single, time-constant $\lambda$ for all constraints, in principle each constraint $j$ could be parameterized with a distinct $\lambda_j$, each of which could also vary as a function of $t$. The number of quantum circuit preparations required to optimize our HTAAC-QSDP protocol is constant with respect to the number of qubits $n$ (and thus to the maximum number of vertices $N=2^n$), as $\langle U_W, \rho_t \rangle$ and $\langle U_P, \rho_t \rangle$ each require only the Pauli-$z$ measurement $\langle \sigma_{n+1}^z \rangle$ on the auxilary qubit, and the $m \sim \text{poly}(n)$ amplitude constraint terms can be calculated from a single set of $n$-qubit measurements on the state $|\psi\rangle$. The classical complexity of each training step scales as just $m \sim \text{poly}(n)$, as one marginal expectation value is calculated from the $|\psi\rangle$ measurements for each of the $m$ constraints.

\subsection{Retrieving SDP Solutions}

\label{sec.solution}

At the end of our protocol, the SDP solution is encoded into $|\psi \rangle$. Like in other QSDP protocols, $|\psi\rangle$ may either be used to extract the full $N$-variable solution or for less computationally intensive analysis (i.e., to characterize the features of the solution or as an input state for further quantum protocols). For many SDPs, such as MaxCut, a good approximation of the solution (here, cut value) can also be extracted from the expectation values comprising the cost function.

To extract this approximate cut value, we note that Eq.\ \ref{eq.maxcut_combinatorial} can be rewritten as

\begin{equation}
    \frac{W_\text{sum}}{2} - \frac{1}{2} \text{min} (\sum_{j<i} W_{ij} v_i v_j),
    \label{eq.maxcut_modification}
\end{equation}

\noindent where $W_\text{sum} = \sum_{j<i} W_{ij}$. We note that this latter sum can be computed on a quantum device as

\begin{equation}
    W_\text{sum} = \frac{N}{2} \langle 0| H^{\otimes n} W H^{\otimes n}|0\rangle,
\end{equation}

\noindent where $H$ is the Hadamard gate, as $H^{\otimes n}|0\rangle$ is the positive-phase and equal superposition of all states. Thus, $W_\text{sum}$ can be estimated with a Hadamard test where $U_W$ is applied to the input state, which we here denote $\langle \sigma_{n+1} \rangle_W^{\otimes H}$. Likewise, note that the second summation in Eq.\ \ref{eq.maxcut_modification} is given by

\begin{equation}
    \sum_{j<i} W_{ij} v_i v_j = \frac{N}{2} \langle \psi| W |\psi \rangle,
\end{equation}

\noindent in the case that the constraints are well-enforced. This allows us to estimate the cut value with $\langle \sigma_{n+1} \rangle_W$, the Hadamard test using the variationally prepared $|\psi \rangle$. Thus, the cost function can be estimated directly as

\begin{equation}
    \langle C_Q \rangle = \frac{N}{4 \alpha} (\langle \sigma_{n+1} \rangle_W^{\otimes H} - \langle \sigma_{n+1} \rangle_W).
    \label{eq.observable_cutvalue}
\end{equation}

If the full solution $|\psi \rangle$ is desired, then full real-space tomography of $|\psi \rangle$ must be conducted by calculating the $N$ marginal distributions of all $k \leq n$ Pauli strings along the $z$ and $x$-axes \cite{D'Ariano2003}, although partial and approximate methods could make valuable contributions \cite{Bisio2009,Kaznady2009}. We now show that once $|\psi \rangle$ is determined, it suffices to assign the partition of each vertex as $v_i = \text{sign}(\psi_i)$, or the sign of the state component $\psi_i$.

In classical semidefinite programming algorithms, such as the Goemans-Williamson algorithm \cite{Goemans1995}, the optimal solution $X^*$ is factorized by Cholesky decomposition into the product $X^* = T^\dagger T$, where $T$ is an upper diagonal matrix. The sign of each vertex $v_i$ is then designated as 

\begin{equation}
    v_i=
    \begin{cases}
      1, & \text{if}\ \mathbf{t}_i \cdot \mathbf{g} \geq 0 \\
      -1, & \text{otherwise},
    \end{cases}
    \label{eq.vertex}
\end{equation}

\noindent where $\mathbf{t}_i$ are the column vectors of $T$ and $\mathbf{g}$ is a length-$N$ vector of normally distributed random variables $g_j \sim \mathcal{N}(0, 1)$.

We define the quantum parallel by noting that as $\rho=|\psi\rangle \langle \psi|$, its Cholesky decomposition is simply the $2^n \times 2^n$ matrix that has the first row $\langle \psi |$ and and all other entries equal to zero. In this decomposition, Eq.\ \ref{eq.vertex} reduces to

\begin{equation}
    v_i=
    \begin{cases}
      1, & \text{if}\ \psi_i \times g_0 \geq 0 \\
      -1, & \text{otherwise}.
    \end{cases}
    \label{eq.classification}
\end{equation}

\noindent As MaxCut has $\mathbb{Z}_2$ symmetry, the cut values are symmetric under inversion, or flipping the sign of all vertices. This makes the sign of the normally distributed $g_0$ irrelevant to the graph partitioning. Without loss of generality, we can therefore set $g_0 = 1$ and classify each vertex as $v_i = \text{sign}(\psi_i)$.

\subsection{Error Convergence}

We now demonstrate that our method has an approximately $O(1/t)$ error convergence rate. Assuming that our loss function $\mathcal{L}$ is a convex function, which is generally only locally and/or approximately satisfied, it has been established that it will converge as $O(1/t)$ for $t$ iterations if it is $L$-Lipschitz continuous \cite{Pena2017}.

While parameterized quantum circuits are known to be $L$-Lipschitz continuous \cite{Patel2021,McClean2018}, we here, in the name or thoroughness, explicitly demonstrate this for our particular loss function. Specifically, a function $f(x)$ is $L$-Lipschitz continuous if

\begin{equation}
    |\nabla f(x) - \nabla f(y)| \leq L|x-y|
\end{equation}

\noindent for all parameter inputs $x$ and $y$. $\mathcal{L}$ will be $L$-Lipschitz continuous if each of its components are independently so.

Starting with the objective function $\langle \sigma_{n+1}^z \rangle_W = \langle \psi| U_W | \psi \rangle$, note that

\begin{equation}
    \partial \langle \sigma_{n+1}^z \rangle_W / \partial \theta_k = i \langle 0 | U_-^\dagger [V_k, U_+^\dagger U_W U_+] U_- |0 \rangle,
    \label{eq.commutator}
\end{equation}

\noindent where $V_k$ is the generator of the unitary matrix parameterized by $\theta_k$. Eq.\ \ref{eq.commutator} is now composed of two terms, each comprised entirely of normal matrices and vectors, save for perhaps the Hermitian generators $V_k$, with some extremal eigenvalue $a_k$. Then, $|\partial \langle \sigma_{n+1}^z \rangle_W / \partial \theta_k| \leq 2 a_k$ and $|\nabla_k \langle \sigma_{n+1}^z \rangle_W - \nabla_k \langle \sigma_{n+1}^z \rangle_W| \leq 4 a_k$. This proof doubles for the constraint terms, save that we replace the unitary matrix $U_W$ with some Pauli string observable.

Fig.\ \ref{fig.6} (left) demonstrates this approximate $O(1/t)$ convergence for the cut value of G11. The approximate nature of this convergence stems from both the discontinuous rounding process and the nonconvexities of the optimization space.

\subsection{Extensions to Other SDPs}

\begin{table*}[htbp]
\caption{\label{tab.table2}
MaxCut statistics for the $800$-vertex graphs G11 (toroid), G14 (skew binary), and G20 (skew integer) for different orders of Pauli string constraints $k$. We compare the best cut value $\text{max}(C_{SDP})$ produced by the leading classical method \cite{Choi2000} compared to those produced by HTAAC-QSDP, with each entry providing the ratio $\text{max}(C_Q)/\text{max}(C_\text{SDP})$ ($\text{mean}(C_Q)/\text{max}(C_\text{SDP})$). With relatively few Pauli string constraints ($k = 4$), our method exceeds the performance of classical methods on all graphs studied.
}
\centering
%\begin{ruledtabular}
\begin{tabular*}{1\textwidth}{@{\extracolsep{\fill}} ccccccc}
\toprule
\textbf{Graph}&
\textbf{$k=2$}&
\textbf{$k=4$}&
\textbf{$k=6$}&
\textbf{$k=8$}&
\textbf{$k=10$}&
\\
%\midrule
G11 & $0.967$ ($0.940$) & $1.019$ ($0.984$) & $1.007$ ($0.999$) & $1.011$ ($0.995$) & $1.022$ ($0.998$)\\
G14 & $1.011$ ($1.000$) & $1.021$ ($1.009$) & $1.023$ ($1.010$) & $1.022$ ($1.014$) & $1.023$ ($1.012$)\\
G20 & $1.007$ ($0.983$) & $1.025$ ($0.993$) & $1.032$ ($0.992$) & $1.049$ ($1.000$) & $1.043$ ($0.993$)
%\bottomrule
\end{tabular*}
%\end{ruledtabular}
\end{table*}

As explained above, the Goemans-Williamson algorithm \cite{Goemans1995} can be applied to numerous other optimization algorithms, such as MaxSat and Max Directed Cut \cite{Goemans1995}. Moreover, HTAAC-QSDP can be adapted to accommodate the constraints of various other SDPs. The use of our method is particularly advantageous when the constraints of a problem are amenable to being expressed through a tractable number of Pauli strings, or when these constraints can be approximately enforced by such a set of strings. The precise mapping of constraints to limited sets of Pauli strings depends on the problem at hand and may require some engineering. We here provide a few such examples.

\subsubsection{Max and Min Bisection}

As one example, consider the Min/Max Bisection problems \cite{Frieze1997}. Min/Max Bisection are particularly relevant to very-large-scale integration (VLSI) for integrated circuit design \cite{Thompson1980}, a vital application area for large-scale SDPs.

The SDPs for estimating the Max Bisection problem has the standard form:

\begin{equation}
\begin{split}
    &\text{minimize}_{X \in \mathbb{S}^+} \hspace{0.2cm} \langle W, X \rangle \\
    &\text{subject to} \hspace{0.2cm} \sum_{i,j} X_{ij} \leq -N/2, \\
    &\text{and} \hspace{0.2cm} X_{ii} = 1, \hspace{0.2cm} \forall i \leq N.
    \label{eq.bisection}
\end{split}
\end{equation}

\noindent The first constraint is equivalent to requiring that half of the variables of $X$ be partitioned equally, hence the term ``bisection''. In analogy with Eq.~\ref{eq.quantum_maxcut}, Eq.~\ref{eq.bisection} can be written as

\begin{equation}
\begin{split}
    &\text{minimize} \hspace{0.2cm} \langle W, \rho \rangle \\
    &\text{subject to} \hspace{0.2cm} \sum_{i,j} \rho_{ij} \leq -N/2 \\
    &\text{and} \hspace{0.2cm} \rho_{ii} = 2^{-n}, \hspace{0.2cm} \forall i \leq N.
    \label{eq.quantum_bisection}
\end{split}
\end{equation}

The second of these two constraints can be enforced by the Pauli strings constraints of Eq.~\ref{eq.constraints}. For large $N$ and assuming no systematic correlations between the ordering of the vertices, the first constraint can be ensured by adding any single Pauli string constraint

\begin{equation}
\langle \mathcal{O}_x \rangle = 0,
\label{eq.constraint_bisection}
\end{equation}

\noindent where $\mathcal{O}_x$ is any Pauli string of $\sigma^x$ operators. To see how Eq.~\ref{eq.constraint_bisection} enforces the first constraint of Eq.~\ref{eq.quantum_maxcut}, consider that any operator $\mathcal{O}_x$ induces a bit-flip on a subset of qubits, such that each state $\psi_i$ is mapped to another state $\psi_{i'}$. This means that $\langle \mathcal{O}_x \rangle = \langle \psi| \mathcal{O}_x | \psi \rangle$ is the sum of $2^{n-1}$ products $2 \psi_i^* \psi_{i'}$, where for each $i$, $|\psi_i| \approx 2^{-n/2}$, as enforced by the Pauli-Z constraints of Eq.~\ref{eq.constraints}. If the probability that a random state $\psi_i$ of $|\psi\rangle$ is positive is $p$, then in the limit of large $N$ and uncorrelated vertex assignment

\begin{equation}
\langle \mathcal{O}_x \rangle = p^2 + (1-p)^2 -2p(1-p).
\label{eq.constraint_bisection_probability}
\end{equation}

The above yields $\langle \mathcal{O}_x \rangle = 0$ \textit{iff} $p=1/2$, which would correspond to the equal partitioning of the vertices required by the Bisection problems. In the case of correlated vertex encodings, the average of several Pauli-$X$ strings $\langle \mathcal{O}_x \rangle$ can be considered. We note that Eq.~\ref{eq.constraint_bisection} can be modified to enforce any partition ratio by solving for $\langle \mathcal{O}_x \rangle$ (Eq.~\ref{eq.constraint_bisection_probability}) with the desired $p$.

\subsubsection{MaxSat}

MaxSat problems are another branch of optimization tasks with constraints that focus on equally weighted vertices. In Max $k$-Sat problems, the number of logical boolean strings of length $k$ are maximized over a given set of boolean variables $v_i$ \cite{Li2021}. For example, Max $2$-Sat is given by \cite{Goemans1995}

\begin{equation}
\begin{split}
    &\text{maximize  } \sum_{j < i} \left[a_{ij} (1-v_i v_j) + b_{ij} (1+v_i v_j) \right] \\
    &\text{subject to  } v_i = \pm 1,
\end{split}
\end{equation}

\noindent where $a_{ij}$ and $b_{ij}$ are the coefficients of the problem. To convert this problem into an SDP, we optimize the objective function

\begin{equation}
\begin{split}
    &\text{minimize}_{X \in \mathbb{S}^+} \hspace{0.2cm} \langle W^-, X \rangle \\
    &\text{subject to} \hspace{0.2cm} X_{ii} = 1, \hspace{0.2cm} \forall i \leq N,
    \label{eq.max2sat}
\end{split}
\end{equation}

\noindent where $W^-_{ij} = a_{ij} - b_{ij}$. Note that Eq.\ \ref{eq.max2sat} is equal to Eq.\ \ref{eq.maxcut} and that the number of satisfied boolean strings can be extracted from an expectation value like Eq.\ \ref{eq.observable_cutvalue}, save that it is now paired with $W_\text{sum} = \sum_{j<i} W^+_{ij}$, where $W^+_{ij} = a_{ij} + b_{ij}$.

\subsubsection{Correlation Matrix Calculation}

Correlation matrices are key to a wide array of statistical applications and can be estimated with limited information using SDPs \cite{Higham2002,Fushiki2009}. In particular, autocorrelation dictates that correlation matrices $X$ have unit diagonals ($X_{ii} = 1, \hspace{0.2cm} \forall i \leq N$), much like MaxCut, Min/Max Bisection, and MaxSat, and can thus be addressed with the rescaled quantum version ($\rho_{ii} = 2^{-n}$, Eq.\ \ref{eq.quantum_norm}) and approximated by the $z$-axis Pauli string constraints. Meanwhile, the extremization of certain correlations (e.g., maximize/minimize $X_{ij}$) and the enforcement of inequality and equality constraints (e.g., $0.2<X_{ij}<0.4$ or $X_{ij}=0.4$) can be enforced by additional constraints with either Pauli strings or the tomography of a select few state components.

\subsection{HTAAC-QSDP with Mixed Quantum States}

\label{sec.mixed_quantum_states}

We now overview the implementation of the HTAAC-QSDP Goemans-Williamson algorithm with mixed quantum states, such as might occur on a noisy quantum device or by interacting the utilized qubits with a set of unmeasured qubits. While the formalism for measuring Pauli strings on such systems is well known, we demonstrate how the required Hadamard Test remains viable. We start with $\rho_R$, some mixed state equivalent of the variational state $|\psi \rangle$. Upon application of the controlled-$U_W$ (without loss of generality for $U_P$) conditioned on the $n{+}1$th qubit, we obtain the density matrix

\begin{equation}
\begin{split}
    &\frac{1}{2} (|0\rangle \langle 0| \otimes I \rho_R I + i |0\rangle \langle 1| \otimes \rho_R U_W^\dagger \\
    -i & |1\rangle \langle 0| \otimes U_W \rho_R I + |1\rangle \langle 1| \otimes U_W \rho U_W^\dagger ).
\end{split}
\end{equation}

\noindent Upon applying a Hadamard gate on the $n{+}1$th qubit and measuring it along the $z$-axis, we obtain

\begin{equation}
    \frac{i}{2}\text{Tr}[\rho_R U_W^\dagger - U_W \rho_R] = \text{Im}[\text{Tr}[U_W \rho_R ]],
\end{equation}

\noindent which is proportional to Eq.\ \ref{eq.objective} with the same coefficient $\alpha$ as prescribed by Eq.\ \ref{eq.unitary_objective}. This enables not only optimization, but also evaluation of the cut value estimation $\langle C_Q \rangle$ (Eq.\ \ref{eq.observable_cutvalue}), which is an accurate representation of the true cut value (Fig.\ \ref{fig.cut_vs_loss}). Alternatively, the element-wise rounded cut value calculated using Eq.\ \ref{eq.classification} can still be utilized as an approximation to the traditional Goemans-Williamson rounding scheme. Although it would not result in the typical inner-product rounding, the sign of each vertex could still defined as the relative sign between each $\psi_i$ and $\psi_0$.

This mixed state formalism of the HTAAC-QSDP Goemans-Williamson algorithm works not only in principle, but also in practice. Fig.\ \ref{fig.cut_vs_epochs} displays the best cut value of a mixed state formalism with otherwise equal parameters (dark blue). The mixed state was generated by adding an unmeasured qubit to the randomly parameterized quantum circuit, which was then traced over prior to minimization and cut classification. The higher rank states moderately improved the performance on most graphs, with a mean higher-rank SDP value of $0.96$ (G11), $1.01$ (G14), and $0.98$ (G20), compared to $0.94$, $1.00$, and $0.98$ in the rank-1 case.

\section{Simulations and Results}

\label{sec.results}

\begin{figure}[t!]
\centering
\includegraphics[width=1\columnwidth]{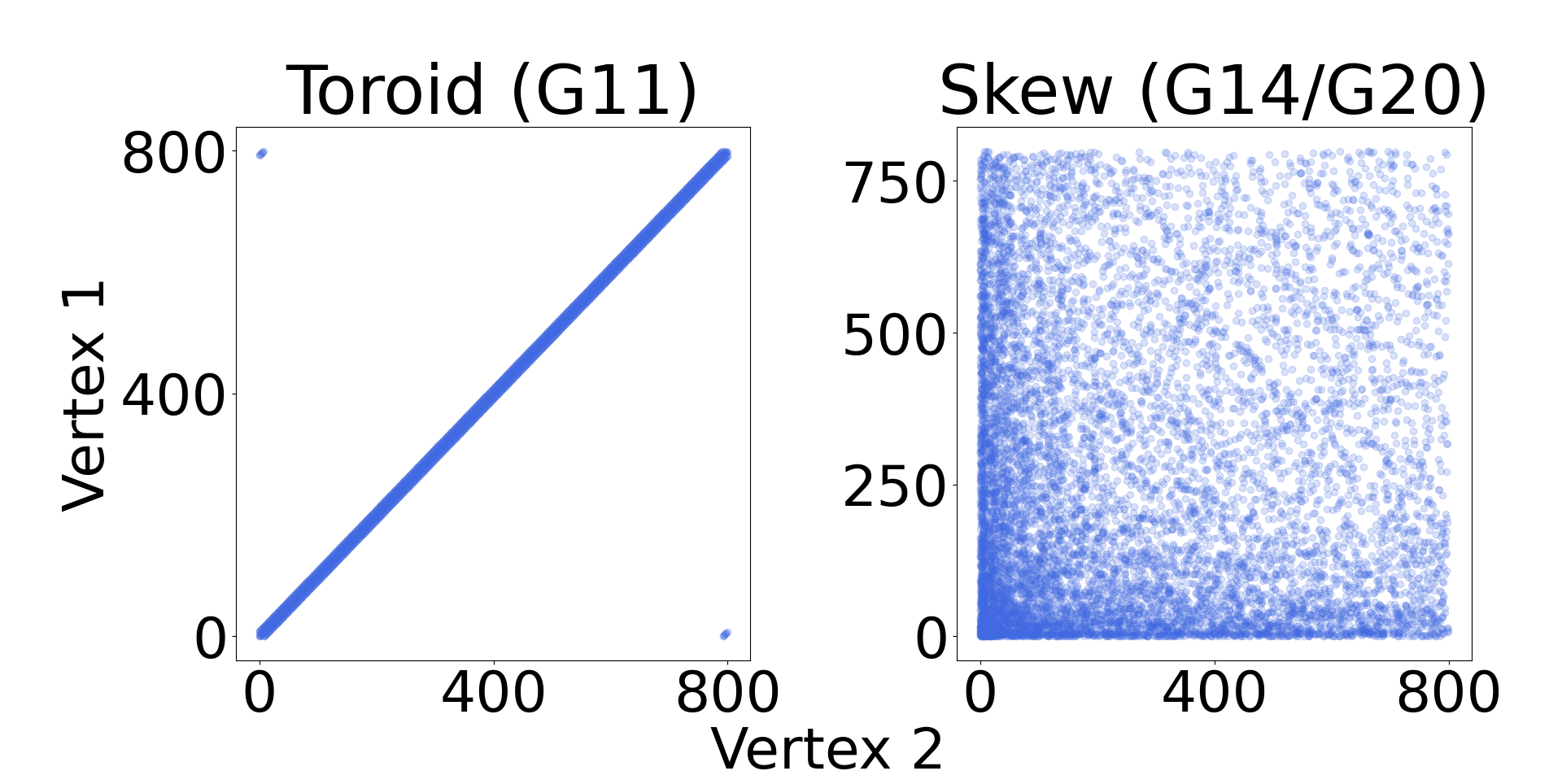}
\caption{The structure of the G11 and G14/G20 GSet graphs, where non-zero edges between two vertices are marked as blue dots. Left) the edges of the toroid graphs (G11, G12, G13) follow a fixed connectivity, with edges extending between neighboring vertices on a torus structure. Right) the skewed graphs have connectivity that is drawn from a random distribution, with edges extending between arbitrary vertices (G14 and G15 binary weights $1$ and $0$, G20 and G21 integer weights $\pm 1$). The degrees of each vertex are disproportionately biased towards the vertices of lower index, with edge density decaying as vertex number increases. We compare with all $800$-vertex graphs considered in the leading classical analog \cite{Choi2000}.}
\label{fig.graphs}
\end{figure}

The viability of our HTAAC-QSDP Goemans-Williamson algorithm is displayed in Fig. \ref{fig.cut_vs_epochs} and Table~\ref{tab.table1}. We compare $C_Q$, the cut values obtained by HTAAC-QSDP, to $\text{max}(C_{SDP})$, the best results obtained by the leading gradient-based classical method \cite{Choi2000}. We study all of the $800$-vertex MaxCut problems explored in \cite{Choi2000} (Table~\ref{tab.table1}) in order to make an extensive comparison with the leading classical gradient-based interior point method. These graphs represent a broad sampling from the well-studied GSet graph library \cite{Benson1999}. Graphs G11, G12, and G13 have vertices that are connected to nearest-neighbors on a toroid structure and $\pm 1$ weights (Fig~\ref{fig.graphs}, right), while G14 and G15 (G20 and G21) have binary weights $0$ and $1$ (integer weights $\pm 1$) and randomly distributed edge density that is highly skewed towards the lower numbered vertices (Fig~\ref{fig.graphs}, left).

HTAAC-QSDP with $k \leq 2$-Pauli term constraints exceeds the performance of its classical counterpart on skewed binary and skewed integer graphs, and falls narrowly short of classical performance on toroid graphs (Fig.~\ref{fig.cut_vs_epochs} and Table~\ref{tab.table1}). The slight differences between the HTAAC-QSDP solution quality and those of the classical solver are typical of comparing different SDP solvers, which often differ slightly in their answers due to different numerical factors, including sparsity tolerance, rounding conventions (especially in the context of degenerate SDP solutions), and other hyperparameters \cite{Todd1999}. All trajectories converge above the $0.878$-approximation ratio $C_Q/C_\text{max}$ (dashed red line) guaranteed by classical semidefinite programming, where $C_\text{max}$ is the highest known cut of each graph found by intensive, multi-shot, classical heuristics \cite{Toshiba}. As SDPs are approximations of the optimization problem, the extremization of the loss function and the figure of merit (here, cut value) are highly correlated, particularly for well-enforced constraints. Fig.~\ref{fig.cut_vs_loss} demonstrates the strong correlation between the cut values estimated by quantum observables $\langle C_Q \rangle$ (Eq.\ \ref{eq.observable_cutvalue}) and the fully rounded SDP result $C_Q$, indicating that the HTAAC-QSDP Goemans-Williamson algorithm measurement of few quantum observables closely approximates the rounded and composited cut values of all variables.

The addition of Pauli string amplitude constraints with $k > 2$ can better enforce Eq.~\ref{eq.quantum_norm}, leading to higher-quality solutions to the SDPs. Fig.~\ref{fig.sigmarhoii_vs_epochs} and Table~\ref{tab.table2} demonstrate that increasing $k$ produces moderately higher $C_Q$ values for the $800$-vertex graphs, until the performance increases saturate $k \approx 4$. Moreover, we note that at $k$ values $\approx 4$, HTAAC-QSDP outperforms the analogous classical algorithm for all graph types, indicating a well-conditioned solution. Likewise, the population variance (solid lines) $\sigma_\rho = \text{var}(|\psi_i|^2)$ decreases substantially until saturating near $k \approx 4$ at around $\sigma_\rho \approx N^{-1}$.

Note that, in the absence of the competing objective function ($\langle \sigma_{n+1}^z \rangle_W$) dynamics, all Pauli-$z$ correlations become restricted as $k \rightarrow n$. This results in the complete constraint $|\psi_i|=N^{-1/2}, \hspace{0.2cm} \forall i$, such that $\sigma_\rho \rightarrow 0$ (black dashed line in Fig.~\ref{fig.sigmarhoii_vs_epochs}, left). We can understand this behavior in the context of other penalty methods \cite{Fletcher1983}. In particular, consider the penalty methods formulation $q(t, \lambda)=f(x_t)+\lambda_t g(x_t)$, where in our case $x_t$ are the time-dependent optimized parameters, $f(x_t)=\langle \sigma_{n+1}^z \rangle_{W,t}$, and $g(x_t)$ represents the constraints. In the limits $t \rightarrow \infty$ and $\lambda_t \rightarrow \infty$, it is known that $x_t \rightarrow \overline{x}$, where $\overline{x}$ is the fully enforced solution of the hard constraints in some neighborhood of $x_0$ \cite{Freund2004}. That is, the constraint-only dynamics (black dashed line of Fig.~\ref{fig.sigmarhoii_vs_epochs}) represents the solution quality with respect to constraints in the $\lambda_t \rightarrow \infty$ limit.

The HTAAC-QSDP Goemans-Williamson algorithm was also tested against the G81 graph, a 20,000 vertex graph with a similar structure to G11-G13, and which is the largest MaxCut graph available that has been benchmarked against classical SDP methods. The average ratio between the HTAAC-QSDP obtained cut value for G81 and that of the classical counterpart was within $1\%$ of the $800$-vertex toroid graphs tested (ratio of $0.93$). While the requisite circuit-depth for some quantum objective functions is known to grow linearly with the size of the circuit's Hilbert space \cite{Anschuetz2021}, this has not been shown for objective functions of the form of Eq.\ \ref{eq.loss}. This is in agreement with our observation that, while Hilbert space required to solve the G81 graphs is $32$ times larger than that for G11-G21, it's circuit-depth is $<8$ times larger. This reduction in requisite circuit-depth could also be due to the influence of constraints, which may effectively limit the degrees of freedom of the quantum circuit, a phenomenon that has been referred to as a ``Hamiltonian informed'' model \cite{Anschuetz2021}.

That being said, even if the worst-case exponential bounds for quantum optimization were saturated, that would not preclude them from having lesser overhead than classical SDPs. We note that classically solving an NP-hard problem of $2^n$ variables is exponentially hard with respect to the $2^n$ variables (that is, it scales as $O(2^{2^n})$) and that, while these problems can be classically approximated using polynomial time and memory with respect to the number of variables, this still yields a classical complexity $\sim \textrm{polynomial}(2^n)$. For instance, the leading classical techniques, interior point methods \cite{Todd1999}, have a memory cost that scales as $O(N^4) \sim O(2^{4n})$ and a computational cost that scales as $O(N^2) \sim O(2^{2n})$, making them in fact less efficient than the exponential bounds put on some worst-case quantum objective functions of $O(2^n/2) = O(2^{n-1}) \sim O(2^n)$ memory cost and $O(2^{n-1}) \sim O(2^n)$ computational cost, a quartic and quadratic reduction, respectively.

Even among the lighter-weight yet often less powerful first-order methods, the classical overhead is considerable. For example, the popular branch of projection-based methods has arithmetic and memory complexity of $O(2^{3n})$ and $O(2^{2n})$, respectively, for such $2^n$ variable problems \cite{Beck2017}. Alternative algorithms, such as the Arora-Kale algorithm, can have time complexity as efficient as $\tilde{O}(M)$ \cite{Arora2016}, although for problems such as MaxCut this still reduces to $\tilde{O}(2^n)$). On a similar topic of scale, we note that G81 was optimized with constraints of orders $k\leq4$, which results in a similar ratio of approximate to exact constraints as that of smaller experiments.

\begin{figure}[t!]
\centering
\includegraphics[width=1\columnwidth]{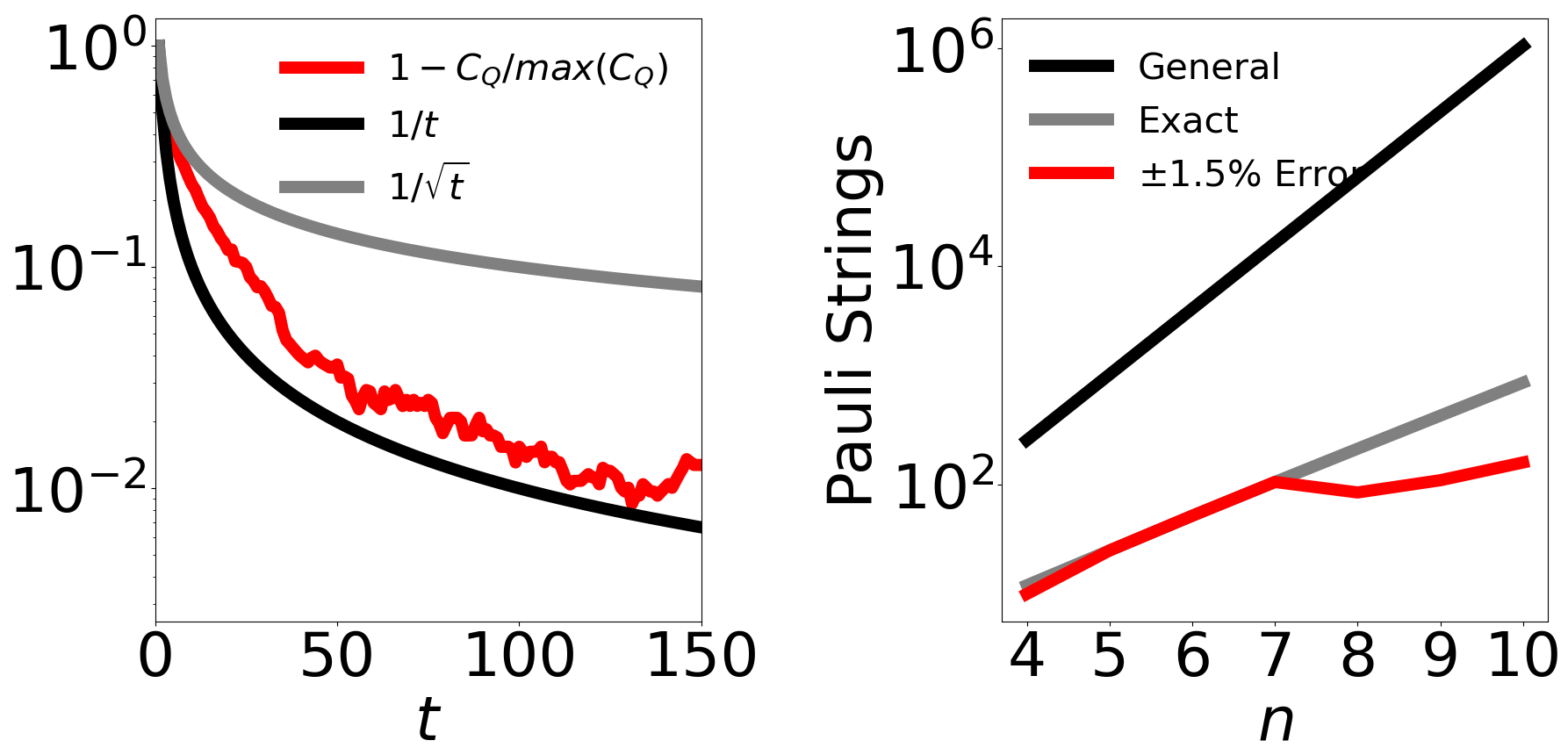}
\caption{(Left) The convergence of the cut value for the G11 graph. The cut value approaches its maximum trajectory value as $\sim 1/t$, with some fluctuations due to both the discontinuity of the rounding process and the nonconvexities of the optimization space. The error convergence is markedly faster than $1/\sqrt{t}$. (Right) Although the general Hilbert space of quantum operators requires $4^n$ operators to span (black), in practice, many SDPs of interest, such as toroid graphs, can be encoded with a polynomial number of Pauli strings (gray). Moreover, many of these Pauli terms are relatively insignificant to the problem structure, such that the approximate matrix $W_\text{approx}$ can be encoded with little error using only a fraction of the already reduced operator set (red).}
\label{fig.6}
\end{figure}

\subsection{Simulation Details}
\label{sec.simulation}

All simulations are done using a one-dimensional ring qubit connectivity, such that each qubit has two neighbors and the $n$th qubit neighbors the $1$st qubit. The circuit ansatz of the simulations for graphs G11-G21 (G81) is $120$ ($900$) repetitions of two variationally parameterized $y$-axis rotations interleaved with CNOT gates, alternating between odd-even and even-odd qubit control. The TensorLy-Quantum simulator~\cite{Patti2021TLQ,tensorly} is used for graphs G11-G21\footnote{Source code at \url{https://github.com/tensorly/quantum/blob/main/doc/source/examples/htaacqsdp.py}.}, while a modified version of cuStateVec \cite{custatevec} is used for G81. Gradient descent was conducted an ADAM \cite{kingma2014adam} optimizer, with learning rate $\eta=0.01$ ($\eta=0.005$) for graphs G11-G21 (G81), as well as hyperparameters $\beta_1=0.9$, and $\beta_2=0.999$.

The evolution angle $\alpha$ was set as $\alpha=0.01$ for all graphs. The values of $\beta$ used in this work were $\beta=1/1.2$ for the toroid graphs and $\beta=1/3$ for the skew binary and skew integer graphs. We did not employ a unitary $U_P$ for the G81 graph. $\beta$ values should be chosen such that $\beta<1$, as diagonal entries are always satisfiable (i.e., some population can always be placed on the state, lowering the loss function), in contrast to edge cuts, which are not (i.e., not every edge can be cut with any given partition for general graphs). $\beta$ values can be tuned on the device in the real time by monitoring the Pauli string constraints and choosing a $\beta$ that leads to largely satisfied Pauli constraints with relatively small coefficients $\lambda$, such that the convergence of the algorithm is not hindered by large constraints that outweigh the objective function or lead to unstable convergence. In this work, we set $\lambda \propto \alpha / m$, to keep the total influence of $m$ constraint terms in proportion to the objective term $\langle \sigma_{n+1}^z \rangle_W \approx \alpha W$. Specifically, for the $800$-vertex graphs, we choose $\lambda = 100 \alpha / m$ for the toroid and skew binary graphs and $\lambda = 50 \alpha / m$ for the skew integer graphs. For the G81 graph, $\lambda = 2000 \alpha / m$.

\section{Theoretical Analysis of Hadamard Test Unitaries}

This subsection addresses the implementation of the Hadamard tests found in this work, including the approximation of $W$ by $U_W$ with a finite phase $\alpha$, the construction of $W$ for difficult problems, and the implementation of the prescribed Hadamard Tests using ZX-calculus.

\subsection{Finite Phase $\alpha$ for Unitary Objective Function}
\label{sec.scale}

In this subsection, we derive Theorem~\ref{theorem.general_claims}, which we here restate for completeness: \\

\noindent \textbf{Theorem 1} \textit{Our approximate Hadamard Test objective function $U_W \sim i \alpha W$ (Sec.~\ref{sec.hadamard}) holds for graphs with randomly distributed edges if}

$$\alpha^2 \lesssim \frac{N^4}{e^3} = \frac{N}{\xi^3},$$

\noindent \textit{where $e$ is the number of non-zero edge weights and $\xi$ is the average number of edges per vertex.} \\

As discussed above, Theorem~\ref{theorem.general_bound} can be understood in two ways: that $\alpha$ satisfies the approximation of Eq.\ \ref{eq.criteria} while remaining tractably large for SDPs of arbitrary $N$, as long as $N$ does not 1) grow slower than the total number of edges $e$, or 2) grow slower than the the cube of $\xi$. We again note that Theorem~\ref{theorem.general_bound} should hold for the densest graph region if the edge density is assymetrically distributed, i.e., $\xi$ should be the average number of edges for the densest vertices. As the conditions of Theorem~\ref{theorem.general_bound} hold for graphs that are not too dense, they are widely satisfiable as the majority of interesting and demonstrably difficult graphs for MaxCut are relatively sparse \cite{Commander2009,Williamson2011,Benson1999,Wiegele2007,DIMACS}. Many classes of graphs for which MaxCut is NP-hard satisfy Theorem~\ref{theorem.general_bound} with tractably large $\alpha$, even for arbitrarily large $N$.

As an example, we consider non-planar graphs, for which optimization problems like MaxCut are typically NP-complete. While planar graphs can be solved in polynomial time \cite{Chaourar2017}, a graph is guaranteeably non-planar when $e > 3N-6$, which reduces to $\xi \geq 3$ in the limit of large $N$ \cite{Makarychev1997}\footnote{Other families of easy graphs are even more restrictive, such as graphs that lack a giant component. In the limit of large $N$, these graphs only occur in more than a negligible fraction of all possible graphs when $d \geq 1/N$ and thus $\xi \geq 1/2$ \cite{Bollobas2001}.}. In accordance with Theorem~\ref{theorem.general_bound}, constant values of $\xi$ actually permit $\alpha$ to \textit{grow} as $N^{1/2}$, while for constant $\alpha$ $\xi$ can grow as $N^{1/3}$, such that our approximation is valid for a wide variety of large-scale non-planar graphs. Indeed, most standard benchmarking graph sets have a small average number of edges per vertex, e.g., $\xi=3$ \cite{Benson1999,Wiegele2007,DIMACS}, as sparse edge-density is common among graphs with real-world applications. In fact, solving MaxCut with many classes of dense graphs (i.e., graphs with nearly all non-zero edges) is provably less challenging, and therefore less interesting, than with their relatively sparse counterparts \cite{Arora1999}.

We here sketch a brief proof of Theorem~\ref{theorem.general_bound} for Erdös–Rényi random graphs \cite{Durrett2006} with edge weights $W_{ij} \sim \mathcal{U}_{[0,b]}$, where $\mathcal{U}_{[0,b]}$ is the uniform distribution on the interval $[0,b]$. The edge density of a graph is described as $d = e/E$, where $e$ is the number of non-zero edges $e$ and $E=N(N-1)/2$ is the number of total possible edges. We provide a detailed proof of this and other graph types in the Appendix~\ref{appendix.scale}. \\

\noindent \textbf{Proof sketch of Theorem~\ref{theorem.general_bound}:}

\begin{itemize}
    \item The Hadamard Test encoding is a good approximation when $U_W \propto i\alpha W$.
    \item This is satisfied when $\frac{\alpha^3}{3!} |W^3|_{ij} \ll \frac{\alpha}{1!} |W|_{ij}$ for typical edges between vertices $i$,$j$.
    \item The mean of the non-zero elements in $W$ is $\overline{W_{ij}} = b / 2$ \footnote{The mean value of \textit{all} elements of $W$ is $\overline{W_{ij}}' = d b / 2$, however the relevant comparison is between the elements of $W^3$ and the non-zero elements of $W$.}.
    \item Elements $(W^3)_{ij}$ are the sum of $\sim N^2$ terms $W_{ij}W_{jk}W_{kl}$, with expectation value $\overline{W_{ij}W_{jk}W_{kl}} = b^3 d^3 / 8$. That is, the additive error between the Hadamard encoding terms and the matrix elements $W_{ij}$ scales as $b^3 d^3 / 8$.
    \item $\frac{\alpha^3}{3!} \overline{(W^3)_{ij}} \ll \frac{\alpha}{1!} \overline{W_{ij}} \rightarrow \alpha^2 \ll 24/(N^2 b^2 d^3)$.
    \item Substituting $d = 2e/N(N-1) \approx 2e/N^2$ and $\xi = e/N$, we obtain Theorem~\ref{theorem.general_bound}.
\end{itemize}

\subsection{Construction of $W$}
\label{sec.Wdesign}

While some optimization problems of $O(2^n)$ variables may only be represented by graphs $W$ of $O(4^n)$ distinct Pauli strings, we here illustrate that there are many interesting (indeed, NP-hard optimization problems) for which this is not the case. In particular, we focus on the MaxCut problem and discuss toroid and Erdos Renyi random graphs.

Toroid graphs, or graphs that can be embedded on a toroid such that none of the edges connecting vertices cross, have a regular, yet still three-dimensional (non-planar), topological structure \cite{chromatic2008}. While encoding difficult problems, these data sets can often be represented in just $\text{poly}(n)$ Pauli strings, as is the case with the 8-100 tourus family to which G11 pertains (Fig.\ \ref{fig.6}, right, gray). What is more, the number of Pauli strings can be reduced even further by instead constructing an approximate operator $W_\text{approx}$ and permitting a small amount of error $\overline{|W_\text{approx}{-}W|} / \overline{|W|} \leq 0.015$ (Fig.\ \ref{fig.6}, right, red). The population balancing graphs $P$ are a similar subset of structured graphs, whose diagonality renders them expressible with Pauli strings of only $z$-axis and identity gates.

Likewise, we can use similarly few terms to construct Erdös–Rényi random graphs, in which edges between any two vertices are equally likely and occur with probability $p$ \cite{Durrett2006}. As Pauli strings are a spanning set, these same statistics are replicated when such strings are chosen randomly. Moreover, we note that each Pauli string adds $O(2^n)$ edges, such that the graph is rapidly populated.

\subsection{Construction of Controlled Unitaries}

The construction of unitary rotations $U_W$ and $U_P$ follows naturally from ZX calculus \cite{VanDeWetering2020}. Specifically, Pauli Gadgets can be used to generalize unitary rotations from the qubit to which they are applied to multiple qubits through the use of $O(n)$ CNOT gates, one on either size of each qubit and its rotated counterparts in a conjugated ladder scheme \cite{Cowtan2019}. Moreover, rotations along distinct Pauli axes are achieved by conjugating these qubits with $\pi/2$ rotation gates along said axis. The gates are selected to match the terms of $W$ or $P$ and $\alpha$ is the phase applied.

As this method generalizes to all rotations, it can also be paired with the controlled gates required for the Hadamard Test. Specifically, the Pauli rotation gate applied to the auxiliary-adjacent qubit is fashioned as a controlled rotation, with the control conditioned on the auxiliary qubit. Moreover, the small values of $\alpha$ used in this work make the addition of multiple Pauli terms by Trotterization favorable, as the error of this technique is bounded by $\alpha^2/2$ times the spectral norm \cite{Childs2021}.

\section{Conclusion}

The efficient optimization of very large-scale SDPs on variational quantum devices has to the potential to revolutionize their use in operations, computer architecture, and networking applications. In this manuscript, we have introduced HTAAC-QSDP, which uses $n+1$ qubits to approximate SDPs of up to $N=2^n$ variables and $M \sim O(N)$ constraints by taking only a constant number of quantum measurements and a polynomial number of classical calculations per epoch. As we approximately encode the SDP objective function into a unitary operator, the Hadamard Test can be used to optimize arbitrarily large SDPs by estimating a constant number of expectation values. Likewise, we demonstrate that the constraints of many SDPs can also be efficiently enforced with approximate amplitude constraints.

Devising a quantum implementation of the Goemans-Williamson algorithm, we approximately enforce the $M=2^n$ constraints with a population-balancing Hadamard Test and the estimation of as few as $m \sim n^2/2$ Pauli string expectation values. We demonstrate our method on a wide array of graphs from the GSet library \cite{Benson1999}, approaching and often exceeding the performance of the leading gradient-based classical SDP solver on all graphs \cite{Choi2000}. Finally, we note that by increasing the order $k$ of our Pauli string constraints, we improve the accuracy of our results, exceeding the classical performance on all graphs while still estimating only polynomially-many expectation values.

The approximate amplitude constraints of HTAAC-QSDP make it particularly helpful for problems with a large number of constraints $M$. The benefits of using the Hadamard Test objective function depend on the original optimization problem. The optimization matrix of many NP-hard problems can be encoded with controlled-unitaries of polynomially-many Pauli terms, such that the Hadamard Test would be efficient to implement. While such cases could instead be optimized by directly estimating polynomially-many different non-commuting expectation values, use of the Hadamard Test circumvents the need to prepare an ensemble of output states for each Pauli term, eliminating these extra circuit preparations. Conversely, optimizing worst-case objective functions with the Hadamard Test would require controlled-unitaries with up to $O(2^n)$ Pauli terms. While exact implementation of these problems with HTAAC-QSDP on purely gate-based quantum computers would be inefficient, such objective functions could be engineered as the natural time-evolution of quantum devices with rich interactions (e.g., quantum simulators \cite{Britton2012,Bernien2017,Paraoanu2014} in their future iterations), or by approximate means.

Due to the immense importance of SDPs in scientific and industrial optimization, as well as the ongoing efforts to generate effective quantum SDP methods that are often limited by poor scaling in key parameters such as accuracy and problem size, our work provides a variational alternative with tractable overhead. In particular, the largest SDPs solved via classical methods, which required over 500 teraFLOPs on nearly ten-thousand CPUs and GPUs \cite{Fujisawa2012}, could be addressed by our method with just $\sim 20$ qubits.

In future work, the techniques of this manuscript can be extended to additional families of SDPs. For instance, SDPs that extremize operator eigenvalues are a natural application for quantum circuits \cite{Lewis1996}. Similarly, variational quantum linear algebra techniques \cite{Xu2021} can potentially be adapted to enforce the more general constraints

$$\langle A_\mu, X \rangle = b_\mu, \hspace{0.2cm} \forall \mu \leq M$$

\noindent of Eq.~\ref{eq.psd}. In many cases, more general constraints are likewise satisfiable with the Pauli string constraints, as suggested in this work. For instance, when the number of requisite constraints $M$ is much smaller than the number of variables $N$, or, as is the case with our quantum implementation of the Goemans-Williamson algorithm, by enforcing a relatively small subset of the constraints.

\section{Acknowledgments}
At CalTech, A.A. is supported in part by the Bren endowed chair, and Microsoft, Google, Adobe faculty fellowships. S.F.Y. thanks the AFOSR and the NSF (through the CUA PFC and QSense QLCI) for funding. The authors would like to thank Matthew Jones, Robin Alexandra Brown, Eleanor Crane, Alexander Schuckert, Madelyn Cain, Austin Li, Omar Shehab, Antonio Mezzacapo, Mark M. Wilde, and Patrick J. Coles for useful conversations.

\bibliographystyle{quantum}
\bibliography{MyCollectionTrunc}

\appendix
\section{Appendix: Theoretical Analysis of Hadamard
Test Objective Function}
\label{appendix.scale}

We now derive Theorem~\ref{theorem.general_bound} in detail. In order for the efficient encoding $U_W = \exp(i\alpha W) \propto i\alpha W$ to hold, it is sufficient to enforce that the third-order term in Eq.\ \ref{eq.unitary} is substantially smaller than the first-order term. That is

\begin{equation}
    \frac{\alpha^3}{3!} |W^3|_{ij} \ll \frac{\alpha}{1!} |W|_{ij} \rightarrow \frac{\alpha^2}{6} |W^3|_{ij} \ll |W|_{ij}
    \label{eq.criteria}
\end{equation}

\noindent for typical edges between vertices $i$,$j$. By induction, the criterion in Eq.\ \ref{eq.criteria} also guarantees that odd (imaginary) powers ${>}3$ will likewise be smaller than the first order term, and are thus also negligible. While this condition can always be satisfied with an arbitrarily small $\alpha$, we in practice require that $\alpha$ maintain some finite size to avoid unitary rotations with vanishingly small gate times $\tau \propto \alpha$ and imaginary components $\langle \sigma_{n+1}\rangle_W \propto \alpha$. We now demonstrate that this criteria can be met for a wide array of graphs with NP-complete MaxCut optimization complexity. \\

First, we consider Erdös–Rényi random graphs \cite{Durrett2006} with elements $W_{ij} \sim \mathcal{U}_{[0,b]}$, which are uniformly distributed on the interval $[0,b]$. The graphs are said to have edge density $d$, which is the fraction of non-zero edges $e$ over total possible edges $E=N(N-1)/2$. Typical elements $(W^3)_{ij}$ are the sum of $\sim N^2$ terms $W_{ij}W_{jk}W_{kl}$, with expectation value $\overline{W_{ij}W_{jk}W_{kl}} = b^3 d^3 / 8$, such that the matrix elements of $W^3$ have the expectation value $\overline{(W^3)_{ij}} = N^2 b^3 d^3 / 8$. As the mean of the non-zero elements in $W$ is $\overline{W_{ij}} = b / 2$, the criterion of Eq.\ \ref{eq.criteria} becomes

\begin{equation}
    \frac{\alpha^2 N^2 b^3 d^3}{48} \ll \frac{b}{2} \rightarrow \alpha^2 \ll \frac{24}{N^2 b^2 d^3}.
    \label{eq.critera_positive_uniform}
\end{equation}

\noindent We can rewrite this criterion in terms the number of non-zero edges $e$ by noting that graph density $d$ scales as $d = e/E$, where $E = N(N-1)/2 \approx N^2/2$ is the number of non-zero edges possible for an $N$ vertex graph. Likewise, the average number of edges per vertex is then $\xi = e/N$, and Eq.\ \ref{eq.critera_positive_uniform} can be rewritten as

\begin{equation}
    \alpha^2 \ll \frac{3 N^4}{b^2 e^3} = \frac{3 N}{b^2 \xi^3}.
    \label{eq.final_alpha_positive_uniform}
\end{equation}

For graphs where edge density $d$ is not uniformly distributed, the above conditions should hold for the most densely connected vertices of the graph. \\

We briefly illustrate how our approximation holds for a few other classes of graphs. For instance, graphs with elements $W_{ij} \sim \mathcal{U}_{[-b,b]}$ drawn from uniform distributions with both positive and negative components generally require $\alpha$ ranges that are even more permissible (i.e., can be even larger) than those of the positive case, with the criterion of Eq.\ \ref{eq.critera_positive_uniform} serving as a small lower-bound. 
%More specifically, the diagonal elements of $W^2$ are the sum of $\sim Nd$ squared terms $(W_{ij})^2$, which have expectation values $\overline{(W_{ij})^2} = b^2/3$, such that the mean of the diagonal elements of $W^2$ is $\overline{(W^2)_{ii}} = Ndb^2/3$. Due to the self-correlation of signs, these diagonal-terms of $W^2$ are significantly larger than their off-diagonal counterparts in $|W^2|$. This leads to a somewhat tighter, yet still worst-case density $d \rightarrow 1$ upper-bound on the mean $\overline{(W^3)_{j \neq i}} = b \overline{(W^2)_{ii}}$ for the dominant off-diagonal terms of $|W^3|$. Comparing this to $\overbar{|W|} = b / 2$ implies the satisfyability condition

%\begin{equation}
%    \alpha^2 \ll \frac{18}{N d b^2} \rightarrow \alpha^2 \ll \frac{9}{\xi b^2},
%    \label{eq.criteria_positive_negative_unitary}
%\end{equation}

%\noindent such that $\alpha$ can be implemented for arbitrary $N$.

Similar proofs of implementability can also be done for graphs with normally distributed weights $W_{ij} \sim \mathcal{N}(\mu, \sigma^2)$ of mean $\mu$ and variance $\sigma^2$. For the case $\mu \not \ll \sigma$, $\overline{(W^3)_{ij}} = N^2 \mu^3 d^3$ and $\alpha$ need only satisfy

\begin{equation}
    \frac{\alpha^2 N^2 \mu^3 d^3}{6} \ll \mu \rightarrow \alpha^2 \ll \frac{6}{N^2 \mu^2 d^3},
    \label{eq.critera_large_mean_unitary}
\end{equation}

\noindent which requires the same permissive scaling between $N$ and $d$ ($e$ or $\xi$) as the condition Eq. \ref{eq.criteria} (Eq.\ \ref{eq.critera_positive_uniform}) for positive uniform distributions. Likewise, for normal distributions where $\sigma \gg \mu$, Eq.~\ref{eq.critera_large_mean_unitary} with $\mu \rightarrow \sigma$ would be a large upper bound.

%\begin{equation}
%    \alpha^2 \ll \frac{6}{N d \sigma^2} \rightarrow \alpha^2 \ll \frac{3}{\xi \sigma^2},
%\end{equation}

%\noindent which requires the same permissive scaling between $N$ and $d$ as the criterion Eq.\ \ref{eq.criteria_positive_negative_unitary}, for graphs with uniform weights $C_{ij} \sim \mathcal{U}_{[-b,b]}$.

\end{document}